\documentclass[aps,prc,twocolumn,showpacs,groupedaddress]{revtex4}
\usepackage{amsmath}

\usepackage{color} 

\definecolor{LightCyan}{rgb}{0.88,1,1}

\usepackage{multirow}

\usepackage{graphicx}
\usepackage{epstopdf}

\usepackage{fancyhdr}
\usepackage{setspace}

\def\pu1 {$^{241}$Pu$^*$\mbox{ }}

\def\g {$\gamma$}
\def\gray{$\gamma$-ray}
\def\grays{$\gamma$ rays}
\def\eg{$\epsilon_\gamma$}
\def\Gf{$\Gamma_f$}
\def\Gg{$\Gamma_\gamma$}
\def\Ggf{$\Gamma_{\gamma f}$}
\def\ngf{$(n,\gamma f)$}
\def\egf{$\overline{\epsilon}_{\gamma f}$}

\def\nubar{$\overline{\nu}$}
\def\nf{$(n,f)$}

\def\egm{\epsilon_\gamma}

\newcommand{\etal}{{\it et al. }}

\begin{document}

\title{Revisiting the role of the \ngf~process in the low-energy fission of $^{235}$U and $^{239}$Pu}
\author{J.E.~Lynn and P.~Talou}
\affiliation{Nuclear Theory Group, Los Alamos National Laboratory, Los Alamos, NM 87545, USA}
\author{O.~Bouland}
\affiliation{CEA, DEN, DER, SPRC, Physics Studies Laboratory, 13108 Saint-Paul-lez-Durance, France}

\begin{abstract}
The \ngf~process is reviewed in light of modern nuclear reaction calculations in both slow and fast neutron-induced fission reactions on $^{235}$U and $^{239}$Pu. Observed fluctuations of the average prompt fission neutron multiplicity and average total \gray~energy below 100 eV incident neutron energy are interpreted in this framework. The surprisingly large contribution of the M1 transitions to the pre-fission \gray~spectrum of $^{239}$Pu is explained by the dominant fission probabilities of 0$^+$ and $2^+$ transition states, which can only be accessed from compound nucleus states formed by the interaction of $s$-wave neutrons with the target nucleus in its ground state, and decaying through M1 transitions. The impact of an additional low-lying M1 scissors mode in the photon strength function is analyzed. We review experimental evidence for fission fragment mass and kinetic energy fluctuations in the resonance region and their importance in the interpretation of experimental data on prompt neutron data in this region. Finally, calculations are extended to the fast energy range where \ngf\ corrections can account for up to 3\% of the total fission cross section and about 20\% of the capture cross section.
\end{abstract}

\date{\today}

\pacs{24.75.+i, 25.85.-w, 25.85.Ec, 23.20.Lv}

\maketitle

\section{Introduction}

\begin{figure}[ht]
\centerline{\includegraphics[width=0.85\columnwidth]{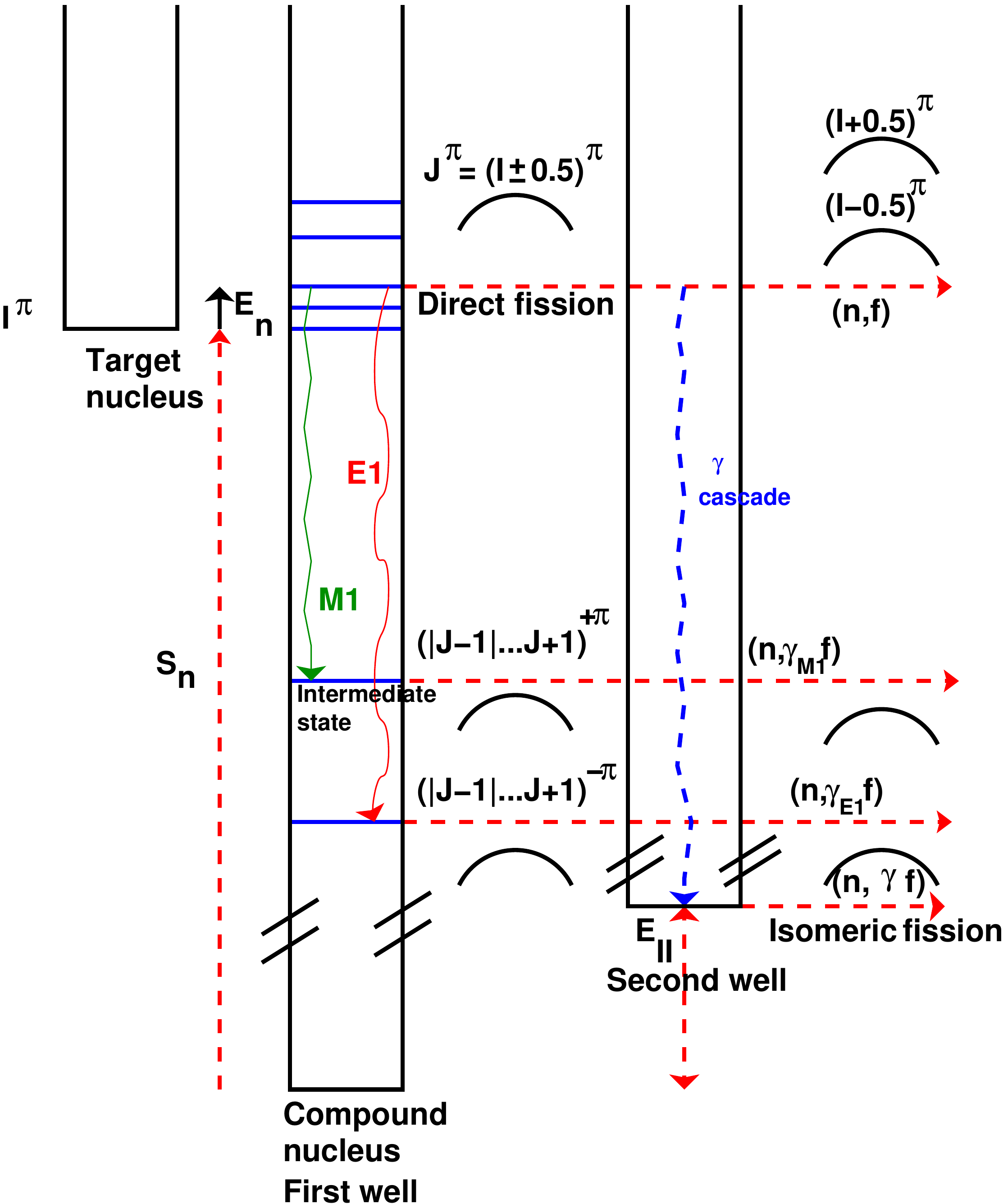}}
\caption{\label{fig:ngf}Two-step ($n,\gamma f$)  processes respectively in first (left-hand side) and second well (right-hand side) of the fission barrier potential energy. I$^\pi$ and J$^\pi$ represent the target and compound nucleus spin and parity, respectively.}
\end{figure}

The decay of a compound nucleus formed by the interaction of low-energy neutrons with a heavy nucleus can happen through neutron emission, radiative capture, or fission. An intriguing scenario occurs when the excited compound nucleus emits a \g~ray but retains enough excitation energy to fission (see Fig.~\ref{fig:ngf}). This \ngf~reaction was first predicted theoretically~\cite{Lynn:1965,Stavinsky:1965} many years ago and calculations made of its magnitude within the framework of a single-humped (i.e., liquid-drop)  barrier. Indirect experimental evidence~\cite{Bowman:1967,Vandenbosch:1967} soon followed, while more direct and compelling evidence came later~\cite{Shackleton:1972,Ryabov:1973,Frehaut:1974,Howe:1976,Moore:1978} through the analysis of neutron-\g~correlations. 

The probability for this process to occur is expected to be small, hence difficult to observe except when the direct fission process is hindered (small fission width) and the $(n,\gamma)$ reaction populates reaction channels more prone to fission. If a \g~ray is emitted prior to fission, slightly less excitation energy is available to the fission fragments, which in turn emit slightly fewer prompt neutrons; this anti-correlation between prompt neutron and \g\ multiplicities has been observed in slow neutron-induced fission reactions on $^{239}$Pu~\cite{Shackleton:1972,Ryabov:1973,Frehaut:1974} and $^{235}$U~\cite{Frehaut:1974,Howe:1976,Moore:1978}.  A comprehensive review of experimental results on this topic has been written by Shcherbakov~\cite{Shcherbakov:1990}. 

\begin{table}[h]
\def\arraystretch{1.5}
\begin{center}
\footnotesize{
\resizebox{0.85\columnwidth}{!} {\begin{tabular}{|c|c|c|c|c|}
\cline{1-5}
 \multicolumn{2}{|c|}{$s$-wave} &    \multicolumn{1}{c|}{$\left\langle \Gamma_{\gamma f}\right\rangle$}& $\left\langle \Gamma_{f}\right\rangle$ & $\left\langle \Gamma_{\gamma}\right\rangle$ \\ 
  \multicolumn{2}{|c|}{resonances} &  (meV) & (meV) & (meV) \\ 
   \hline
 $^{236}$U$^{*}$  &  $3^-$ & $0.87\pm0.89$~\cite{Shcherbakov:1990} & $180 \pm 18$~\cite{Moore:1978} &  \\  
 \cline{3-3}
&   & $4.7\pm2.3$~\cite{Trochon:1980} &&    \\
 \cline{3-3}
&   & $3.0$~\cite{Lynn:1965} &&  $38.1\pm1.7$~\cite{Mughabghab:2006}  \\ 
 \cline{2-4}
 &  $4^-$ & $0.32\pm0.13$~\cite{Shcherbakov:1990} &  $91 \pm 11$~\cite{Moore:1978} & \\ 
 \cline{3-3}
&   & $\leq 1.2$ & & \\ 
 \cline{3-3}
 &   & $1.5$~\cite{Lynn:1965} & &  \\
 \cline{3-3}
&   & $2.1\pm0.7$~\cite{Trochon:1980} & & \\
 \cline{1-5} \cline{1-5} 
$^{240}$Pu$^{*}$  &  $0^+$ & $2.8\pm9.2$~\cite{Shcherbakov:1990}  &  2270~\cite{Trochon:1970} &   \\ 
 \cline{3-3}
 & & $7.3\pm1.8$~\cite{Trochon:1980} & & \\
 \cline{3-3}
 & & $4-7$~\cite{Lynn:1965} & & \\
  \cline{3-3}
&  & 5.73~\cite{Shackleton:1972} &  &  $43.0\pm4.0$~\cite{Mughabghab:2006} \\ 
 \cline{2-4}
 &  $1^+$ & $1.91\pm0.81$~\cite{Shcherbakov:1990}  &  $33.7\pm5$~\cite{Trochon:1970} & \\ 
  \cline{3-3}
& & 3.0~\cite{Lynn:1965} &    & \\ 
  \cline{3-3}
& & 2.76~\cite{Shackleton:1972} &    & \\ 
 \cline{3-3}
&   & $4.1\pm0.9$~\cite{Ryabov:1973} & &  \\
 \cline{3-3}
&   & $4.2\pm0.4$~\cite{Trochon:1980} & &  \\
  \cline{1-5}
 \end{tabular}}
 }
\caption{Comparison of two-step fission widths $\langle \Gamma_{\gamma f}\rangle$ with evaluated direct fission widths $\langle \Gamma_{f}\rangle$, as reported in the review by Shcherbakov~\cite{Shcherbakov:1990}. For reference,  $s$-wave average capture widths $\langle \Gamma_\gamma\rangle$ from the Atlas~\cite{Mughabghab:2006} are listed as well. References to the original experimental papers are also given.}   
\label{tab:widths}
\end{center}
\end{table}

Table~\ref{tab:widths} summarizes the average widths for the total fission (\Gf), post-\g\ or ``two-step'' fission (\Ggf), and capture (\Gg) processes dominant in the low-energy neutron-induced reactions on $^{235}$U and $^{239}$Pu, as reported by Shcherbakov~\cite{Shcherbakov:1990} and Mughabghab~\cite{Mughabghab:2006}. Those numbers  demonstrate the dominance (on average) of direct fission for $3^-$ resonances in $^{236}$U$^*$ and $0^+$ resonances in $^{240}$Pu$^*$. The large magnitude of the $0^+$ fission will likely mask any experimental investigation of the competitive two-step process for that channel. On the other hand, the small width for the $1^+$ direct fission channel provides a more likely candidate for observing this effect.

\begin{figure}[ht]
\includegraphics[width=\columnwidth]{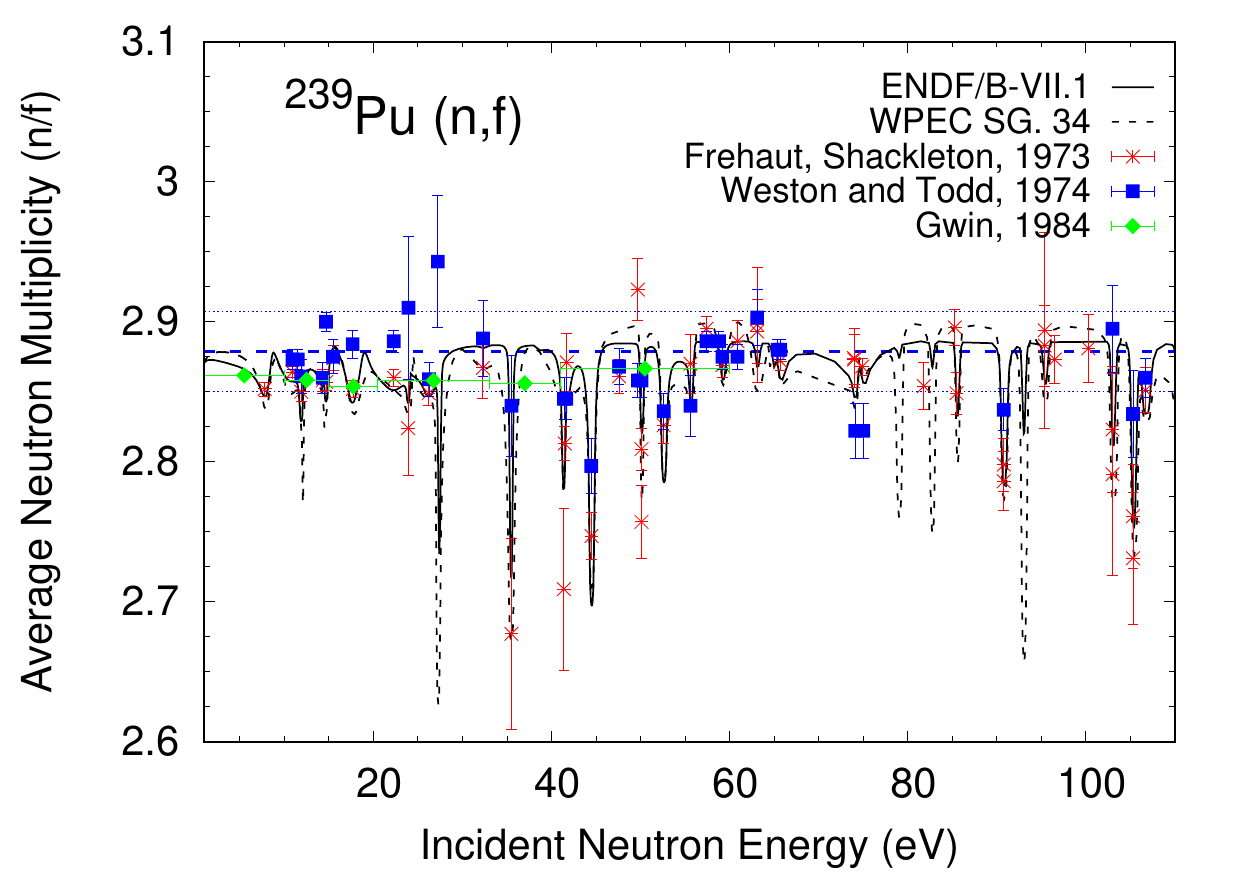}
\caption{\label{fig:nubar-RRR-Pu239} Fluctuations in the average prompt fission neutron multiplicity \nubar\ for the $^{239}$Pu $(n,f)$ reaction. While the evaluated results from ENDF/B-VII~\cite{ENDFB7} and the work by the WPEC Subgroup 34~\cite{WPEC34} clearly show distinct fluctuations in \nubar, the experimental data may not be as convincing or even consistent. More precise experiments on this important quantity should be attempted. The $\pm$1\% error band centered around the evaluated thermal value~\cite{Carlson:2009} is also shown as a guide. Experimental data are from Frehaut \etal~\cite{Shackleton:1972}, Weston and Todd~\cite{Weston:1974}, and Gwin~\etal~\cite{Gwin:1984}.}
\end{figure}

\begin{figure}[ht]
\includegraphics[width=\columnwidth]{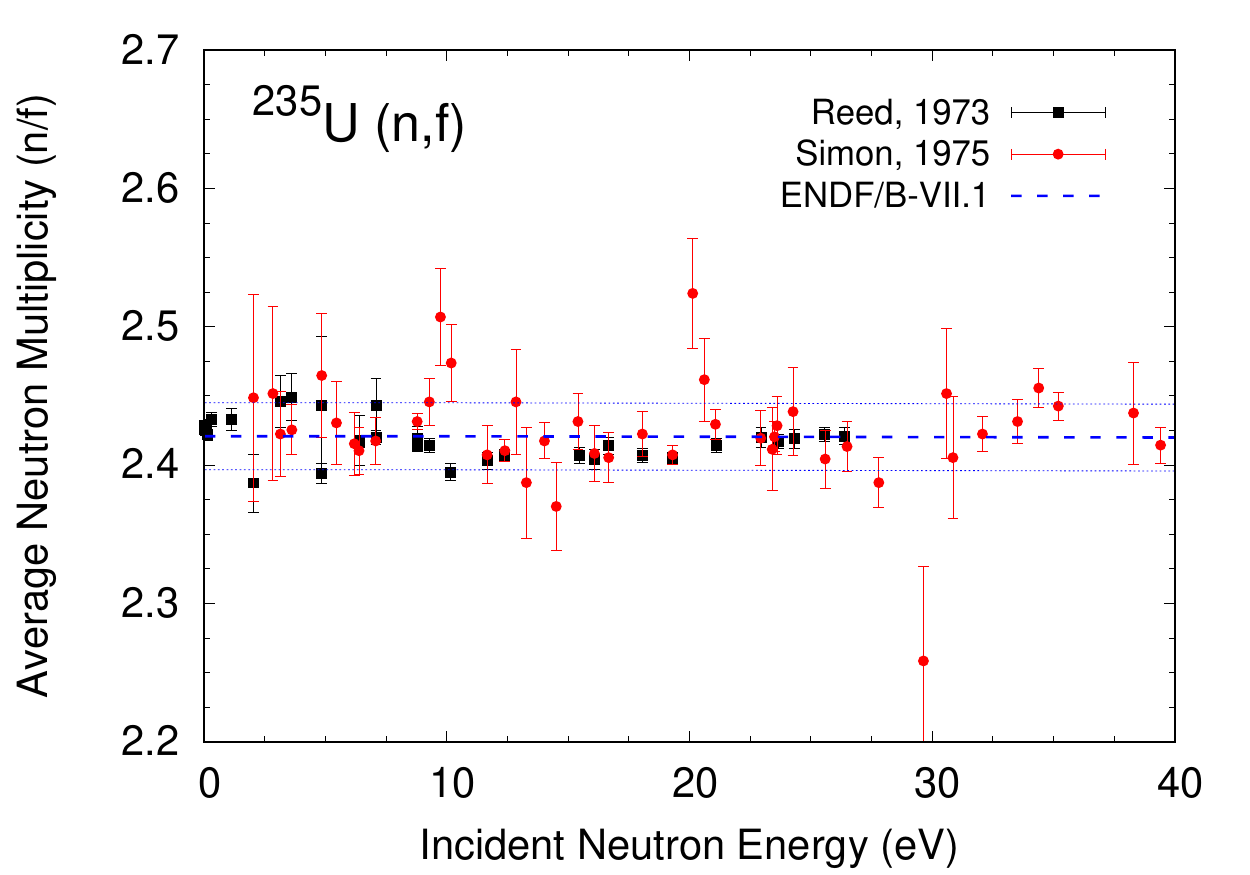}
\caption{\label{fig:nubar-RRR-U235}Same as Fig.~\ref{fig:nubar-RRR-Pu239} for neutron-induced fission on $^{235}$U. The observed fluctuations are relatively small (see 1\% error band in dotted lines for guidance) compared to those observed for $^{239}$Pu. Experimental data are by Reed~\etal~\cite{Reed:1973} and Simon~\etal~\cite{Simon:1975}.} 
\end{figure}

This study is important not only to deepen our theoretical understanding of the fission process, but also to support modern evaluations of nuclear data for applications. Indeed, the fluctuations of the average prompt fission neutron multiplicity \nubar\ in the resonance region of $^{239}$Pu, as shown in Fig.~\ref{fig:nubar-RRR-Pu239}, and evaluated in~\cite{Fort:1988}, were shown to impact nuclear reactor benchmarks. In his review paper, Shcherbakov~\cite{Shcherbakov:1990} emphasizes the anti-correlation observed between the decrease in neutron multiplicity and the increase in \g\ multiplicity. More recently, this evaluation was revisited as part of an international effort to study the slow neutron-induced reactions on $^{239}$Pu~\cite{WPEC34}. The importance of those fluctuations is emphasized in a recent IAEA co-ordinated research project on the prompt fission neutron spectrum of actinides~\cite{Capote:2016}, as prompt fission neutron spectra and multiplicities play important compensating roles in the correct description of criticality benchmarks, for instance. Much smaller fluctuations can perhaps be observed in the resonance region of $^{235}$U \nf\ shown in Fig.~\ref{fig:nubar-RRR-U235}. Fluctuations of \nubar\ have also been interpreted~\cite{Hambsch:1989,Fort:2008,Hambsch:2017} as the result of different fission modes populated at resonance energies. Higher average total kinetic energies, $\langle$TKE$\rangle$, could indeed cause a drop in \nubar, but would not be able to account for corresponding changes in the total prompt \gray\ energy, as discussed in Section~\ref{sec:discussion}.

The study of the pre-fission \gray\ spectrum is invaluable as well, in particular to learn about the nature, electric or magnetic, and multipolarity of the \g\ transitions occurring between highly excited states in the fissioning compound nucleus. The stronger than expected role played by M1 transitions was pointed out in previous experimental analyses (see Ref.~\cite{Shcherbakov:1990} and references therein), and can be explained in the present theoretical study. More recently, experimental and theoretical studies~\cite{Guttormsen:2012,Ullmann:2014,Mumpower:2017} pointed to the existence of a low-lying M1 resonance known as {\it scissors mode}. We investigate the impact of this additional component in the photon strength function on the predicted pre-fission \gray~spectrum.

Obviously, the experimental observation of the pre-fission \gray~spectrum faces significant challenges. First, the average multiplicity of pre-fission \grays~is necessarily small compared to the average multiplicity $\overline{N}_\gamma$ of prompt fission \grays. Second, the relative contribution of the \ngf~process is highest when the fission width \Gf~is small, hence difficult to measure. Finally, because the pre-fission \grays~cannot be easily differentiated from the prompt \grays~emitted by the fragments, their observation can mostly be done using ratio methods in which two spectra, with and without the \ngf~contribution, are measured in similar experimental setups.

In this paper, we perform new calculations of the \ngf~width using our modern understanding of the double-hump fission barriers and resonance \g\ emission. In the picture of a double-humped fission barrier, two-step \ngf\ processes can, in principle, occur in both wells (see Fig.~\ref{fig:ngf}) but it can be shown that the probability of it happening in the second well is much weaker by about two or more orders of magnitude than in the first well. Hence, we will not comment on this second-order correction any further, except to note that it would be important for any calculations of shape isomer yield. 

In addition to calculating \Ggf, we also compute the mean energy \egf\ and spectrum $N_{\gamma f}$ of the \grays\ preceding fission, and correspondingly, the excitation energy loss of the compound nucleus prior to fission. For this purpose we require a knowledge of the level density in spin and parity, radiative capture and fission strength functions. In the following sections, we describe the formulations we use for the \g$f$ width (Section~\ref{sec:formalism-ngf}), level densities (Section~\ref{sec:formalism-LD}), radiative strength functions (Section~\ref{sec:formalism-Gg}) and fission probabilities (Section~\ref{sec:formalism-fission}). Sections~\ref{sec:results} and~\ref{sec:fastresults} summarize our numerical results for the slow and fast neutron-induced reactions on $^{239}$Pu and $^{235}$U, respectively, followed by a discussion in Section~\ref{sec:discussion}. Finally, Section~\ref{sec:conclusion} summarizes our findings and discusses potential extensions to this work.

\section{\label{Sec:Formalism}Formalism and Modeling}

\subsection{Expressions for the \ngf~process}\label{sec:formalism-ngf}

The width \Ggf~for the \ngf~process can be obtained by calculating first the probability that a capture event occurs, and multiplying it by the probability for the residual nucleus, after \g~emission, to fission. In mathematical terms, the width \Ggf~for a compound nucleus with excitation energy $E^*$, spin $J$ and parity $\pi$ reads
\begin{widetext}
\begin{eqnarray} \label{eq:Ggf}
\Gamma_{\gamma f}(E^*,J^\pi) = \sum_{Xl} \sum_{J_f=|J-l|}^{J+l}{\int_0^{E^*}{d\epsilon_\gamma \rho \left(E^*-\epsilon_\gamma,J_f^{\pi(-)^{Xl}} \right)\Gamma_{\gamma Xl}(\epsilon_\gamma)} P_f \left(E^*-\epsilon_\gamma,J_f^{\pi(-)^{Xl}}\right)}, 
\end{eqnarray}
\end{widetext}
where $Xl$ follows the conventional notation for multipolarity $l$ of type $X$ ($E$ or $M)$. The nuclear level density of the fissioning nucleus $\rho$ is considered at the residual excitation energy $(E^*-\epsilon_\gamma)$ after the emission of one \g~ray with energy \eg. The initial excitation energy is $E^*=E_{\rm inc}+B_n$, with $E_{\rm inc}$ the incident neutron energy and $B_n$ the neutron binding energy of the target nucleus. In this work we assume that no more than one \g~ray can be emitted prior to fission. The probability for more than one \g~ray to be emitted prior to fission is most certainly negligible. The capture width \Gg~is calculated for the electric dipole $E1$ and magnetic dipole $M1$ transitions only. Finally, $P_f$ represents the fission probability calculated at the residual excitation energy $E^*-\epsilon_\gamma$, spin $J_f$ and parity $\pi_f$. The term $(-)^{Xl}$ follows the parity conservation rule, as follows
\begin{equation}
\pi(-)^{Xl}= \begin{cases}
\pi, & \text{if $Xl=M1$},\\
-\pi, & \text{if $Xl=E1$}.
\end{cases}
\end{equation}
In this work, $E2$ transitions are neglected. 

The spectrum of the primary \grays~preceding fission can be obtained as
\begin{eqnarray} \label{eq:primaryspec}
N_{\gamma f} (E^*,\epsilon_\gamma) = \frac{d\Gamma_{\gamma f} (E^*)}{d\epsilon_\gamma},
\end{eqnarray}
where \Ggf~is given by Eq.~(\ref{eq:Ggf}). The mean energy $\overline{\epsilon}_{\gamma f}$ of the \g~ray preceding fission, easily obtained by multiplying the integrand in Eq.~(\ref{eq:Ggf}) by $\epsilon_\gamma$.

We now review the calculation of each term entering in Eq.~\ref{eq:Ggf} in the following subsections.

\subsection{Compound Nucleus Level Density} \label{sec:formalism-LD}

We use the QPVR model to generate numerical combinations of Quasi-Particle, Vibrational and Rotational states. The development and motives for using the QPVR level density model, particularly in the context of fission, are given in Ref.~\cite{Bouland:2013}. The energies of the resulting states are calculated by simply summing the energies and axial spin projections of the quasi-particle, vibrational and rotational components. These are placed in bins of 0.1 MeV width and labelled by total angular momentum and parity. The quasi-particle states are generated from the single-particle Nilsson states for neutrons and protons~\cite{Gustafson:1967} as a function of prolate deformation. The deformation parameter we use is 0.25. The Nilsson diagrams give nucleon orbital energies $e_i$ in terms of the mean oscillator frequency $\hbar\omega_0$, where $\omega_0$ represents the spherical circular frequency of the oscillator. A global estimate for this quantity is $\hbar\omega_0=41A^{-1/3}$. The quasi-particle energies $\epsilon_i$ are then calculated from the Nilsson states using
\begin{eqnarray}
\epsilon_i=\sqrt{\left(e_i-e_F\right)^2+\Delta^2},
\end{eqnarray}
where the ``Fermi" energy $e_F$ is taken, for simplicity, to be halfway between the last partially or fully occupied Nilsson orbital and the next unoccupied orbital when filling the Nilsson states with the available number of nucleons.

For $^{240}$Pu$^*$, we use 6.598 MeV for the parameter $\hbar\omega$. The Fermi energies are 6.34 $\hbar\omega$ for protons and 7.446 $\hbar\omega_0$ for neutrons. For the pairing energy parameter $\Delta_p$ for protons we use 0.81 MeV and for neutrons $\Delta_n$=0.56 MeV. For the vibrational states we adopt the $K^\pi$ values and energies of the measured spectrum of $^{240}$Pu. The beta-vibration energy is 0.861 MeV and other collective states are shown in Table~\ref{tab:collective}. The rotational band energy constant is assumed to be 0.0065 MeV. With these parameters the spacing of $J^\pi=1^+$ states at the neutron separation energy is 3.03 eV, in agreement with neutron resonance data~\cite{Mughabghab:2006}. In the $^{236}$U$^*$ calculations most of the parameters are the same with the exception of the pairing gap parameters ($\Delta_p$=0.868 MeV, $\Delta_n$=0.577 MeV) and the collective vibration energies ($E_\beta$=0.919 MeV, $E_{\rm ma}$=0.688 MeV, $E_\gamma$=0.958 MeV).

Discrete inelastic levels relevant to the treatment of the unresolved resonance range are also included in this model.

\subsection{Expressions for Radiation Widths} \label{sec:formalism-Gg}

The model used by Bouland \etal\cite{Bouland:2013} is summarized here. For E1 transitions our \gray~strength function is the sum of a valence term and a Brink giant resonance form with energy-dependent damping width similar to that adopted by Kopecky and Uhl~\cite{Kopecky:1990}:
\begin{eqnarray}
\Gamma_G(\egm) = \Gamma_{G_0}\frac{B\egm^2+A}{E_{G_i}^2},
\end{eqnarray}
in which we use $B=1$ and $A=10$. The E1 strength function for partial radiation widths of \gray~energy \eg~is
\begin{widetext}
\begin{eqnarray} \label{eq:GE1}
\frac{\Gamma_{\gamma E1}(\egm)}{D\egm^3} = 0.418 \times 10^{-9} A^{2/3}+ 4.62\times 10^{-6}\frac{NZ}{N+Z}
\sum_{k=1,2}{\frac{k}{3}\frac{\Gamma_{Gk}(\egm)\egm}{\left( \egm^2-E^2_{Gk}\right)^2+\egm^2\Gamma^2_{Gk}(\egm) }}
\end{eqnarray}
\end{widetext}

Similarly, we use a sum of valence and giant resonance terms for the M1 transitions. The numerical constants of these terms are adjusted to satisfy the evidence on the relative strength of M1 and E1 transitions compiled by Kopecky and Uhl~\cite{Kopecky:1990}. The M1 strength function is 
\begin{widetext}
\begin{eqnarray} \label{eq:GM1}
\frac{\Gamma_{\gamma M1}(\egm)}{D\egm^3} = 0.237 \times 10^{-9} A^{1/3}+ 0.536\times 10^{-7}A^{1/3}
\frac{\Gamma_{GM1}\egm}{\left( \egm^2-E^2_{GM1}\right)^2+\egm^2\Gamma^2_{GM1}}
\end{eqnarray}
\end{widetext}
The M1 giant resonance parameters are set at their values recommended by Kopecky and Uhl, i.e., $E_{\rm GM1}$=6.6 MeV and $\Gamma_{\rm GM1}$=4 MeV.

The total radiative capture width is the sum of the partial radiation widths to all lower states of the compound nucleus. In heavy non-magic nuclides, such as the actinides, it can, with good approximation, be limited to the E1 and M1 transitions and expressed as an integral:
\begin{widetext}
\begin{eqnarray}
\Gamma_{\gamma (tot)}(E^*,J,\pi) = \sum_{J_f=|J-1|}^{J+1}\left\{ \int_0^{E^*}{d\egm \rho \left(E^*-\egm,J_f^{-\pi} \right)\Gamma_{\gamma E1}(\egm)} +
 \int_0^{E^*}{d\egm \rho \left(E^*-\egm,J_f^\pi \right)\Gamma_{\gamma M1}(\egm)} \right\}.
\label{eq:Ggtot} 
\end{eqnarray}
\end{widetext}

Among the actinides, the average radiation width of the $^{240}$Pu neutron cross-section resonances is probably the most accurately known. Mughabghab~\cite{Mughabghab:2006} recommends $\Gamma_{\gamma(tot)}$=31$\pm$2 meV, but several resonances have assigned errors of 1 meV and that of the 1.04 eV resonance has been measured as 30.27$\pm$0.06 meV~\cite{Spencer:1986}. The parameters used in our model, especially the constants associated with the M1 radiative strength function, have been adjusted somewhat to closely reproduce this value. The level density function $\rho(E)$ is obtained from the QPVR model, with pairing gap parameters adjusted to reproduce the resonance spacing of $^{240}$Pu, as described in Section~\ref{sec:formalism-LD}. With these parameters the total radiation width of the $0^+$ resonances in the cross-section of $^{239}$Pu is calculated to be 37.9 meV and that of the $1^+$ resonances to be 38.5 meV, close to the measured values.

\begin{figure}[ht]
\centerline{\includegraphics[width=\columnwidth]{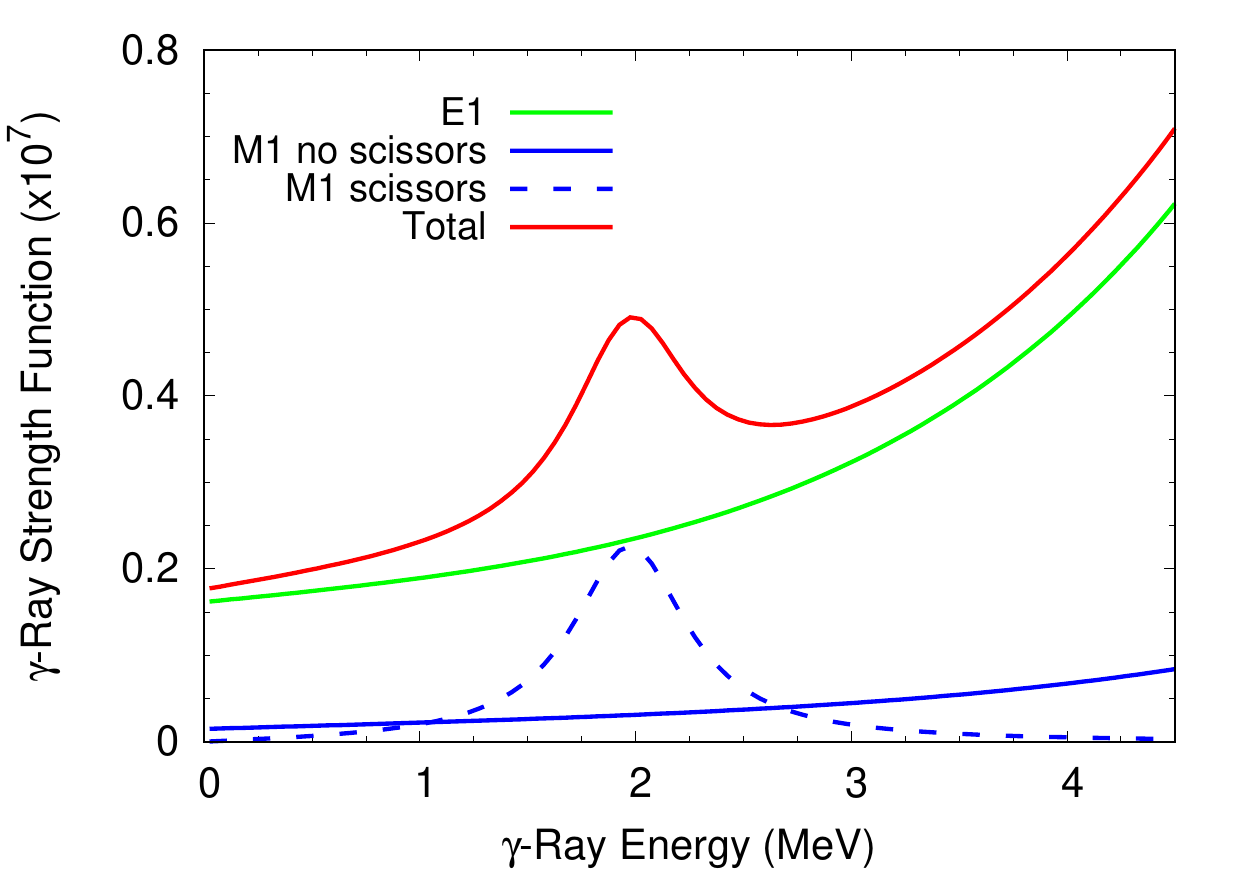}}
\caption{\label{fig:SFcomponents}The E1 photon strength function of Eq.~(\ref{eq:GE1}) (green) and the M1 function of Eq.~(\ref{eq:GM1}) (blue) are shown, as well as the additional M1 scissors resonance in Eq.~(\ref{eq:scissors}) (dashed blue). The total calculated photon strength function including the scissors mode is shown in solid red.}
\end{figure}

There is now much experimental evidence and theoretical support for an additional low-lying M1 resonance known as the scissors mode in the photon strength function. Ullmann \etal~\cite{Ullmann:2014} have recently found that adding a scissors mode resonance, in qualitative agreement with the Oslo data~\cite{Guttormsen:2012,Guttormsen:2014} improves their representation of the shape of the \gray~spectra observed in neutron resonance capture by $^{238}$U.  The energy of this mode appears to be about 3 MeV in the lanthanides and around 2 MeV in the actinides. In Fig.~\ref{fig:SFcomponents}, we show the radiative strength function components of Eqs.~(\ref{eq:GE1}) and (\ref{eq:GM1}) above and a postulated M1 scissors,
\begin{eqnarray}
\frac{\Gamma_{\gamma M1}^{sc}}{D\epsilon_\gamma^3} = 8.67\times 10^{-8}\sigma_{M1,sc}\Gamma_{M1,sc}\frac{\Gamma_{G,sc}\epsilon_\gamma}{(\epsilon_\gamma^2-E_{G,sc}^2)^2+\epsilon_\gamma^2\Gamma_{G,sc}^2},
\end{eqnarray}
in the form used in Ref.~\cite{Mumpower:2017}. The value used in Ref.~\cite{Mumpower:2017} for the mode strength $\sigma_{M1,sc}\Gamma_{M1,sc}$, derived mostly from analysis of capture cross-section data in the fission product region, is $42.2\beta_2^2$, where $\beta_2$ is the nuclear prolate deformation parameter. With a reasonable value assumed for $\beta_2$, this is of the order of 1 to 2 mb.MeV. In the work of Ullmann \etal~\cite{Ullmann:2017} on the capture \gray\ spectra and capture cross-sections of a range of uranium isotopes, the Oslo data were re-analyzed revealing a double hump in the scissors mode (see Fig.~2 of~\cite{Ullmann:2017}). For the mode near 2 MeV the mode strength found using different methods of analysis ranged from 0.32 to 0.84 mb.MeV and $\Gamma_{M1,sc}$ = 0.8 MeV, while for that near 2.9 MeV the values lie between 0.24 and 0.56 mb.MeV. For our calculations of the effect of a possible scissors mode on our radiative strength function, shown in Fig.~\ref{fig:SFcomponents}, we use only the lower mode with a mode strength of 0.34 mb.MeV and  $\Gamma_{M1,sc}$ = 0.65 MeV. The importance of this mode in the $^{239}$Pu\ngf\ reaction is that it enhances fission through $J^\pi=0^+$ and 2$^+$ states reached from the $1^+$ resonances. The upper mode should have little effect on the \ngf\ process because its primary \grays\ will mainly terminate in an excitation range where the fission probability is negligible.

\subsection{Fission Probabilities} \label{sec:formalism-fission}

The energy, spin and parity-dependent fission probabilities $P_f$ appearing in Eq.~(\ref{eq:Ggf}) are calculated in the R-matrix model for fission as described in~\cite{Bjornholm:1980}. This formalism was recently applied successfully across a suite of plutonium isotopes~\cite{Bouland:2013}. This model is particularly suited to the description of low-energy nuclear fission when the excitation energy of the compound nucleus is near or below the fission barrier. The existence of a second minimum along the fission path and the coupling of class-I and class-II states are taken into account by sampling the characteristics of those states (widths, energies and coupling strengths) using the Monte Carlo technique. The coupling between class-I and class-II states modifies the Hauser-Feshbach equations, and its impact on the fission widths below the fission barrier is significant.

Particle-transfer reactions, such as $(t,pf)$~\cite{Cramer:1970}, have been used to infer fission barrier characteristics, including height, width, pairing energies, and level densities at the saddle points. Two models for the intermediate resonances are described below.

\subsubsection{Vibrational resonance model: secondary well vibrations}

We assume in our first model that the ``resonances" observed in the fission probability are due to states of $\beta$-vibrational character in the secondary well of the fission barrier. These are built on top of the shape isomer, i.e., the ``ground-state" of the secondary well, with energy $E_{II,G}$. When admixed into the class-I compound states, these $\beta$-states give the amplitude for crossing the inner and outer barriers, and hence govern the magnitude of the coupling and fission widths for passing through the basic barrier transition states with no intrinsic excitation. Let us denote the phonon energy of the $\beta$-vibration by $\hbar\omega_\beta$. For a pure vibrational state, the fission (outer barrier) and coupling widths are respectively
\begin{eqnarray}
\Gamma_{Vf} &=& P_B\hbar\omega_B/2\pi, \\
\Gamma_{Vc} &=& P_A\hbar\omega_A/2\pi,
\end{eqnarray}
where $P_A$ and $P_B$ are the penetration functions of the inner and outer barriers. They are usually calculated from the Hill-Wheeler formula
\begin{eqnarray} \label{eq:HW}
P_i=\frac{1}{1+\exp{\left(V_i-E\right)/\hbar\omega_i}}.
\end{eqnarray}
We now assume that this state is admixed with the many other configurations that are possible for a compound nucleus with considerable excitation energy into the secondary well compound nucleus states (class-II) with a damping width $\Gamma_{VD}$. The fission strength function for excitation energy $E_{II}=E-E_{II,G}$ in the secondary well is given by
\begin{eqnarray}\label{eq:GammaFission}
\frac{\Gamma_{II,f}}{D_{II}} = \sum_n{\frac{\Gamma_{Vf}\Gamma_{VD}}{\left( n\hbar\omega_\beta-E_{II}\right)^2+\left( \Gamma_{VD}/2\right)^2}},
\end{eqnarray}
where $n$ denotes the number of phonons in the vibrational state $V$. The coupling strength is given by a similar equation
\begin{eqnarray}
\frac{\Gamma_{II,c}}{D_{II}} = \sum_n{\frac{\Gamma_{Vc}\Gamma_{VD}}{\left( n\hbar\omega_\beta-E_{II}\right)^2+\left( \Gamma_{VD}/2\right)^2}}.
\end{eqnarray}
The $\beta$-states can also be coupled to excited intrinsic states, which can be other forms of collective motion (e.g. rotation) or quasi-particle excitation, and thus govern the coupling and fission widths through other barrier transition states. The $\beta$ phonons carry no angular momentum or parity, so the vibrational resonances are characterized by the $K,J$ and parity quantum numbers of the intrinsic state with which the $\beta$-state is coupled. Equation~\ref{eq:GammaFission} for fission through such transition states is generalized to
\begin{eqnarray}
\frac{\Gamma_{II,f_j}}{D_{II}} = \sum_n{\frac{\Gamma_{Vf_j}\Gamma_{VD}}{\left( E^{int}_j + n\hbar\omega_\beta-E_{II}\right)^2+\left( \Gamma_{VD}/2\right)^2}},
\end{eqnarray}
where $E^{int}_j$ is the excitation energy of the intrinsic state at the secondary well deformation. A similar equation holds for the coupling width:
\begin{eqnarray} \label{eq:GammaCoupling}
\frac{\Gamma_{II,c_j}}{D_{II}} = \sum_n{\frac{\Gamma_{Vc_j}\Gamma_{VD}}{\left( E^{int}_j + n\hbar\omega_\beta-E_{II}\right)^2+\left( \Gamma_{VD}/2\right)^2}}.
\end{eqnarray}
The damping width is expected to be quite strongly dependent on excitation energy. We assume an exponential dependence and use
\begin{eqnarray} \label{eq:damping}
\Gamma_{VD} = \Gamma_{VD_0}\exp{\left[ \left( E-E_{II,G} \right)\kappa_D\right]}.
\end{eqnarray}
with $\kappa_D$, denoting the vibrational damping coefficient.\\

We show in Table~\ref{tab:collective} the intrinsic states that we expect to be of significance for slow neutron-induced fission of an even compound nucleus and that are used in our calculations.

\begin{widetext}
\begin{center}
\begin{table}[ht] 
\def\arraystretch{1.25}
\begin{tabular}{cccccccc}
\hline
Character & $K^\pi$ & $J^\pi$ (rotn.) & $E^{\rm int}(\beta_I)$ & $E^{\rm int}(\beta_A)$ & $E^{\rm int}(\beta_{II})$ & $E^{\rm int}(\beta_B)$ & $\Gamma_{VD_0}$ \\
\hline
Nil ('ground') & $0^+$ & $2^+, 4^+, 6^+, ...$ & 0.0 & 0.0 & 0.0 & 0.0 & 0.1 \\
Octupole (mass asym.) & $0^-$ & $1^-, 3^-, 5^-, ...$ & 0.597 & 0.7 & 0.6 & 0.1 & 0.1 \\
Octupole (bending) & $1^-$ & $2^-, 3^-, 4^-, ...$ & 0.94 & 0.8 & 0.65& 0.55 & 0.2 \\
Gamma & $2^+$ & $3^+, 4^+, 5^+, ...$ & 1.14 & 0.15 & 0.8 & 0.8 & 0.2 \\
Mass asym. + bending & $1^+$ & $2^+, 3^+, 4^+, ...$ & 1.56 & 1.15 & 1.35 & 1.0 & 0.3 \\
Mass asym. + gamma & $2^-$ & $3^-, 4^-, 5^-, ...$ & 1.56 & 0.85 & 1.4 & 0.9 & 0.3 \\
2 gamma & $0^+$ & $2^+, 4^+, 6^+, ...$ & & 0.45 & 1.6 & 1.5 & 0.3 \\
2 gamma & $4^+$ & $5^+, 6^+, 7^+, ...$ & & 0.37 & 1.6 & 1.7 & 0.3 \\
2 quasi-particle & $0^-$ & $1^-, 2^-, 3^-, ...$ & & 1.74 & 1.45 & 1.74 & 0.4 \\
\hline
\end{tabular}
\caption{\label{tab:collective}Properties of the collective states assumed in the determination of the level spectra along the fission path. All the energies and widths are given in MeV. Nominal damping widths $\Gamma_{VD_0}$ in the last column refer to the double-hump barrier model. The labels ``A, B" correspond to the first and second barriers, and the labels ``I,II" to the first and second wells.}
\end{table}
\end{center}
\end{widetext}

\begin{figure}[ht]
\centerline{\includegraphics[width=\columnwidth]{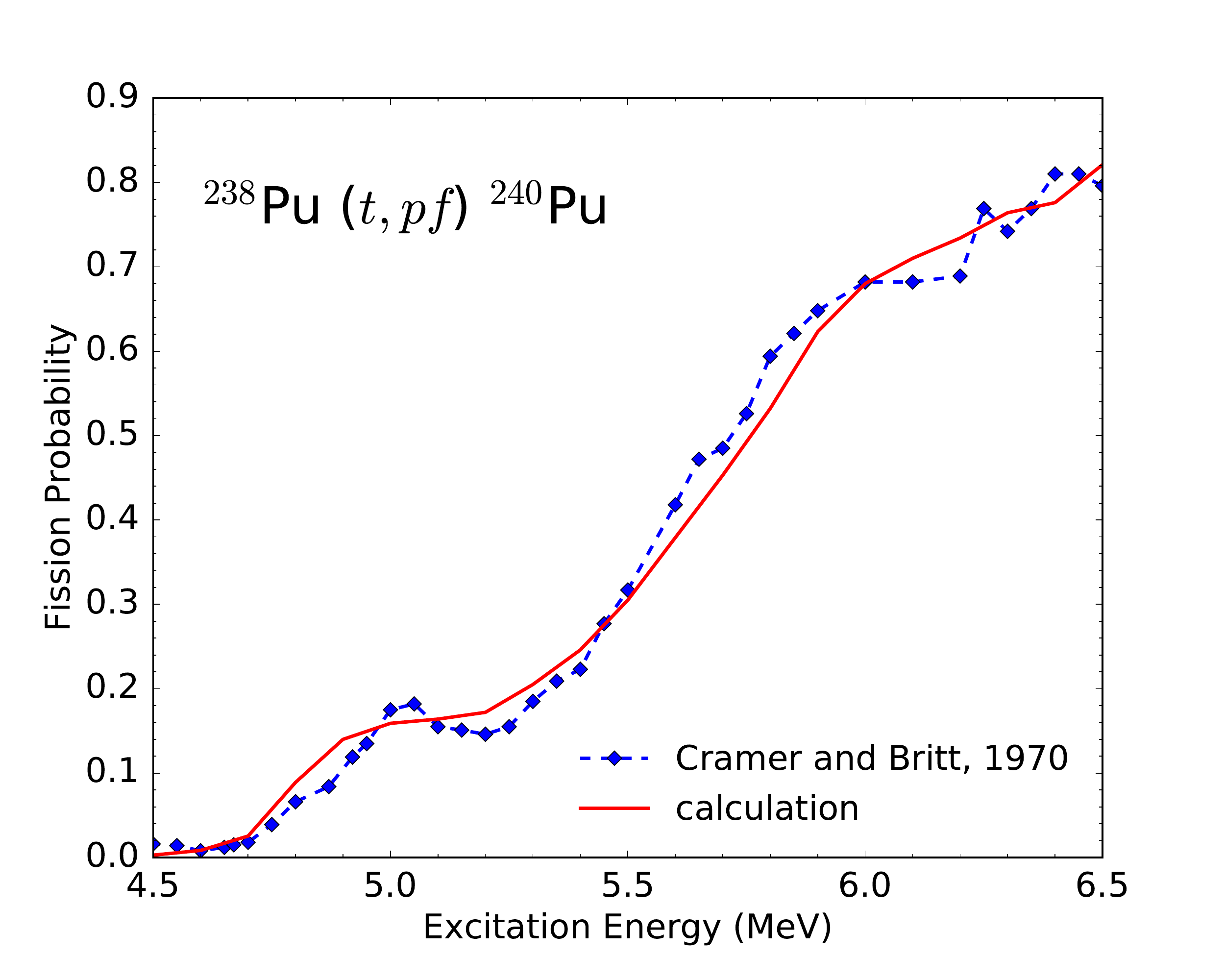}}
\caption{\label{fig:pu238-tpf}Fission probabiliy extracted from the $^{239}$Pu ($t,pf$) transfer reaction. Experimental data are from Cramer and Britt~\cite{Cramer:1970}.}
\end{figure}

\begin{figure}[ht]
\centerline{\includegraphics[width=\columnwidth]{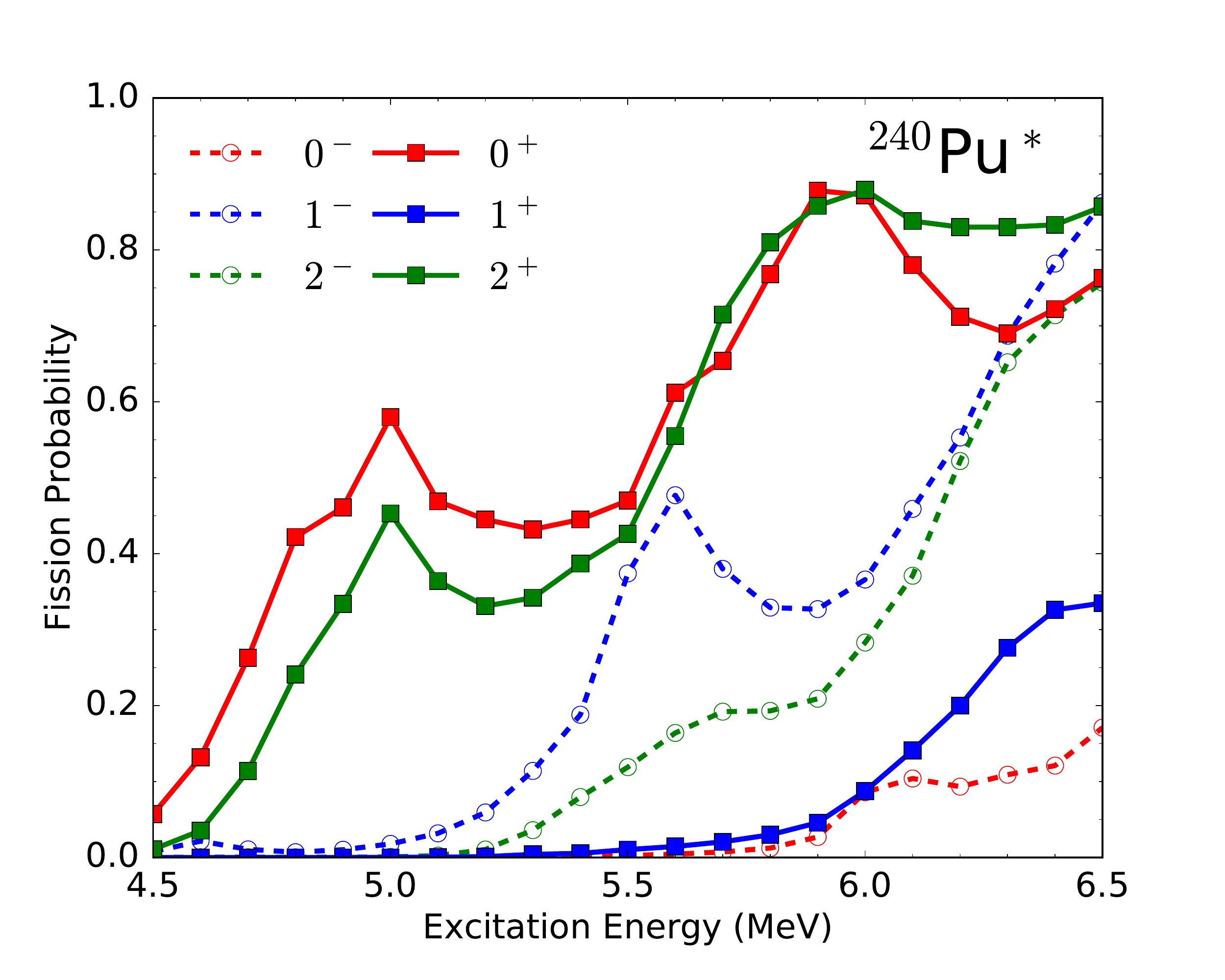}}
\caption{\label{fig:fp-pu240-secondary}Spin and parity-dependent fission probabilities calculated for $^{240}$Pu$^*$, using the secondary vibrational model. Positive-parity states are depicted with full symbols and solid lines, while negative-parity states are shown with empty symbols and dashed lines.}
\end{figure}

The fission probability calculated for $^{240}$Pu$^*$ is shown in Fig.~\ref{fig:pu238-tpf} along with experimental data by Cramer and Britt~\cite{Cramer:1970}, adjusted by a normalization factor of 1.35 to bring them in line with the ratio of neutron capture to fission cross sections in the neutron energy range up to 200 keV, which has been measured to considerable accuracy. This normalization factor is reasonable given the uncertainties in the inference of the fission probabilities from the $(t,pf)$ reactions as well as in the model calculations. Since $^{239}$Pu is fissile, i.e., its highest fission barrier lies below its neutron separation energy, fission barrier information can only be inferred indirectly, e.g., from transfer-induced fission reactions such as the $^{238}$Pu $(t,pf)$ reaction studied by Cramer and Britt~\cite{Cramer:1970}. A reasonable fit to those probabilities as a function of excitation energy $E^*$ could be obtained with the following model input parameters:
\begin{eqnarray}
V_A = 5.65 \mbox{ MeV} &\mbox{ ; }& \hbar \omega_A = 1.05 \mbox{ MeV} \nonumber \\
V_B = 5.23 \mbox{ MeV} &\mbox{ ; }& \hbar \omega_B = 0.6 \mbox{ MeV},
\end{eqnarray}
where $V_A$ and $V_B$ correspond to the inner and outer barrier heights respectively, and $\hbar \omega_A$ and $\hbar \omega_B$ being the corresponding fission barrier widths. Other parameters used in the fit are a secondary well ground-state energy $E_{II,G}$ of 2.95 MeV. This value is based on shape isomer data; the compilation of Singh \etal\cite{Singh:2002} gives 2.8 MeV, but we have adjusted this somewhat to better fit the vibrational resonances, a $\beta$-phonon energy of 1 MeV and rotational band constants $h^2/2\mathcal{I}$ of 3.33 keV at the deformation of the inner barrier, 3.5 keV at the secondary well and 2.5 keV at the outer barrier. The damping coefficient $\kappa_D$ of Eq.~\ref{eq:damping} is chosen to be 0.1 MeV$^{-1}$. In Fig.~\ref{fig:fp-pu240-secondary}, we show the fission probabilities calculated using the same model for states of spin and parity that can be reached by E1 and M1 emissions from the $s$-wave resonances in the neutron-induced reactions on $^{239}$Pu.

\subsubsection{\label{ss:ter}Tertiary well vibration model}

\begin{figure}[ht]
\centerline{\includegraphics[width=\columnwidth]{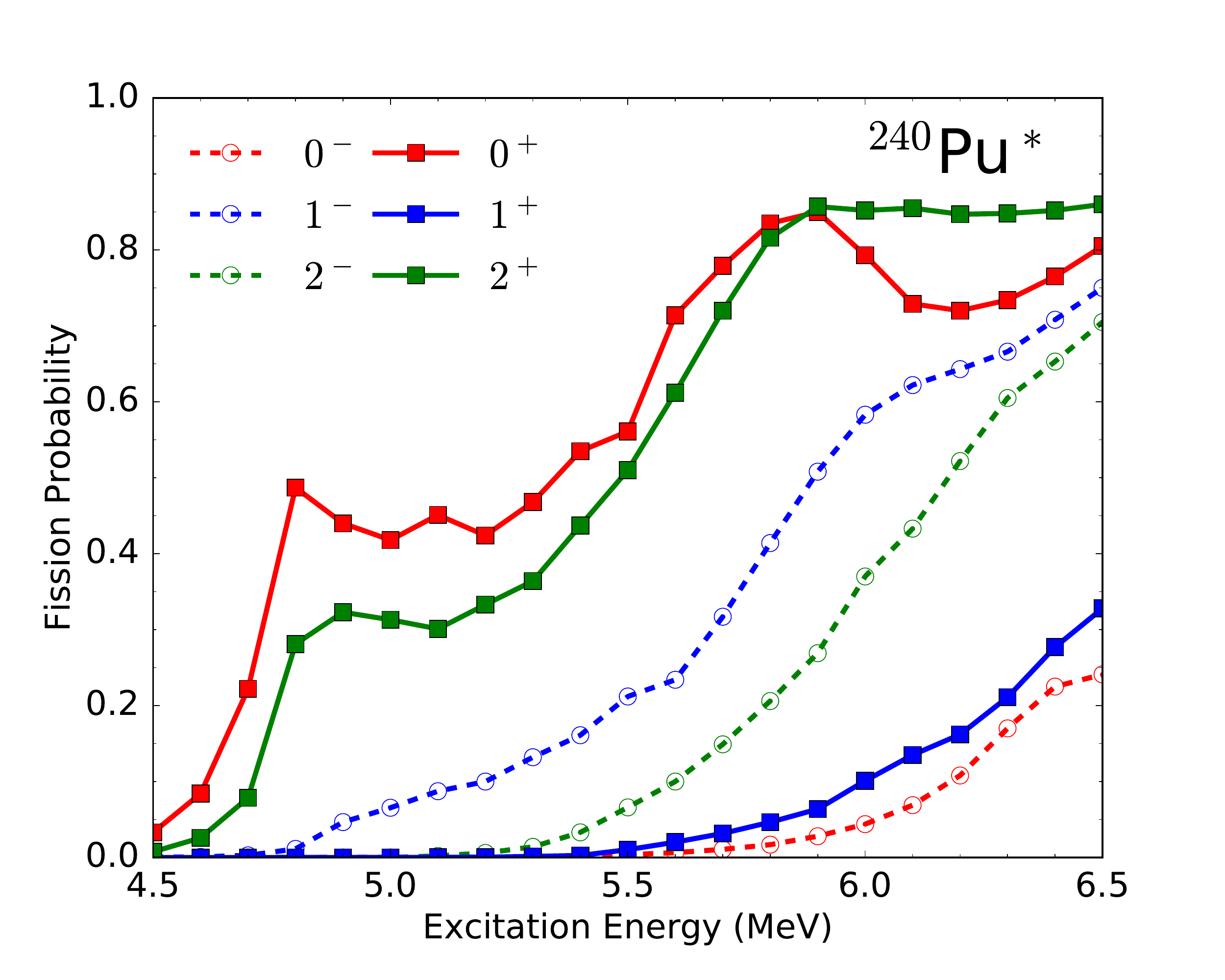}}
\caption{\label{fig:fp-pu240-tertiary}Same as Fig.~\ref{fig:fp-pu240-secondary} but in the tertiary well vibrational model.}
\end{figure}

An alternative assumption is that the vibrational resonances are associated with vibrations within a shallow tertiary well in the general deformation region of the outer barrier~\cite{Ichikawa:2013}. Although the damping of these tertiary well vibrations is probably very small, we assume values of 30 keV and 50 keV for the two lowest states in Table~\ref{tab:collective}.  We can model this approximately by replacing Eq.~(\ref{eq:GammaCoupling}) for the coupling width by the ``strong" coupling assumption embodied in the Hill-Wheeler formula of Eq.~(\ref{eq:HW}):
\begin{eqnarray}
\frac{2\pi\Gamma_{II,c}}{D_{II}} = P_A = \frac{1}{1+\exp{\left[ \left( V_A-E\right)/h\omega_A\right]}}.
\end{eqnarray}
We then assume for $^{240}$Pu$^*$ a shallow tertiary well and place its ``ground" state at 4.8 MeV. The $\beta$-phonon energy is assumed to be 1 MeV, while the damping coefficient, $\kappa_D$, is assumed to be 1 MeV$^{-1}$. Figure~\ref{fig:fp-pu240-tertiary} shows the revised spin and parity-dependent fission probabilities obtained in this model.

\subsubsection{$n+^{235}$U}

\begin{figure}[ht]
\centerline{\includegraphics[width=\columnwidth]{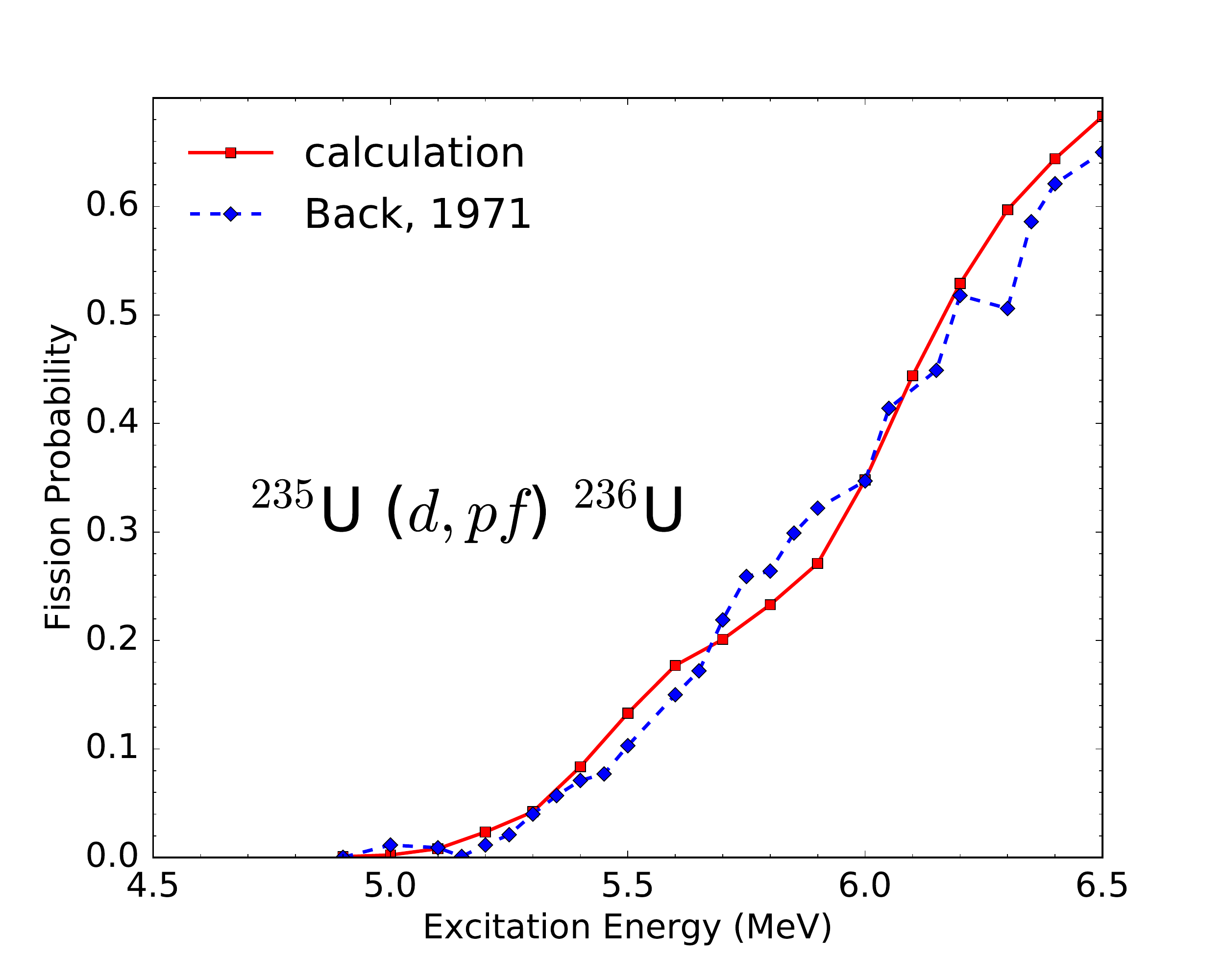}}
\caption{\label{fig:u235-dpf}Fission probability extracted from the $^{235}$U$(d,pf)^{236}$U reaction, with experimental data by Back~\cite{Back:1971}.}
\end{figure}

We have made similar analyses for $^{236}$U$^*$. In this case, most weight was placed on the $^{235}$U($d,pf$) reaction as measured by Back \etal\cite{Back:1971}. A range of possible pairs of values of the inner and outer barrier heights was found. In accordance with a range of theoretical potential landscape studies that suggest that the inner and outer barriers are about equal in the uranium isotopes, we use the values $V_A$=5.56 MeV, $\hbar \omega_A$=1.05 MeV, $V_B$=5.56 MeV, $\hbar \omega_B$=0.6 MeV, and a similar shallow tertiary well vibrational resonance model to that described above for $^{240}$Pu but with a higher ``ground" state energy of $5.37$ MeV. The corresponding fit to the $(d,pf)$ data is shown in Fig.~\ref{fig:u235-dpf}.

\begin{figure}[ht]
\centerline{\includegraphics[width=\columnwidth]{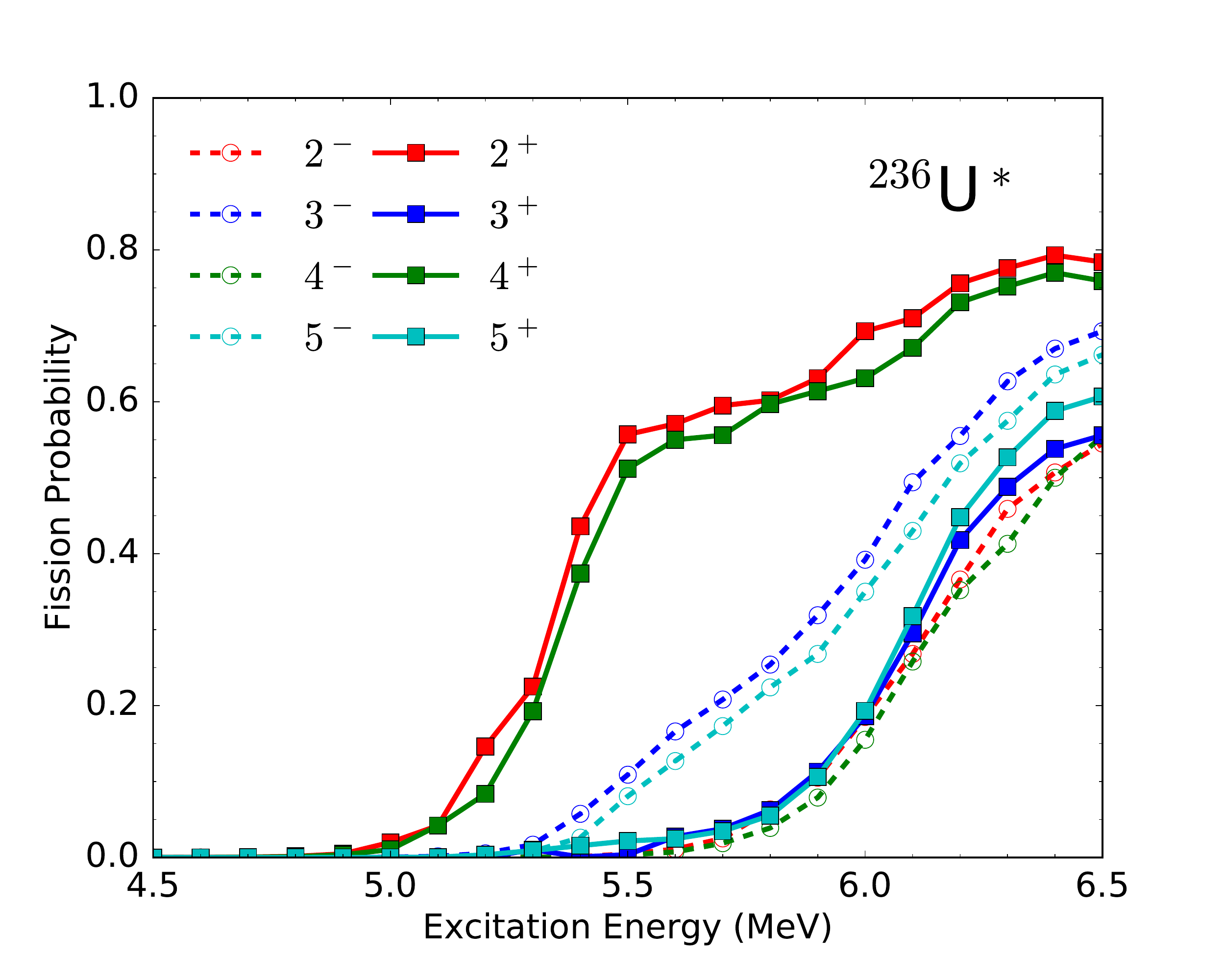}}
\caption{\label{fig:fp-u236}Spin and parity fission probabilities calculated for $^{236}$U$^{*}$, in the tertiary well assumption.}
\end{figure}

The fission probability calculations with the tertiary well model are shown in Fig.~\ref{fig:fp-u236}, indicating that at low excitation energies, the fission probabilities are dominated by the $2^+$ and $3^+$ states.

\section{Slow neutron energy resonances} \label{sec:results}

\subsection{$1^+$ Resonances of $^{239}$Pu + $n$}

With our radiation and level density models and adopted parameters the total radiation width of the 1$^+$ resonances is calculated to be 38.5 meV (without the scissors mode).  This is to be compared with the average, 37.5 meV, of the parameters listed in ENDF/B-VII.1 for the resonances up to 20 eV. Mughabghab~\cite{Mughabghab:2006} recommends a value of 43$\pm$4 meV from earlier available data and analyses. 

\begin{figure}[ht]
\centerline{\includegraphics[width=\columnwidth]{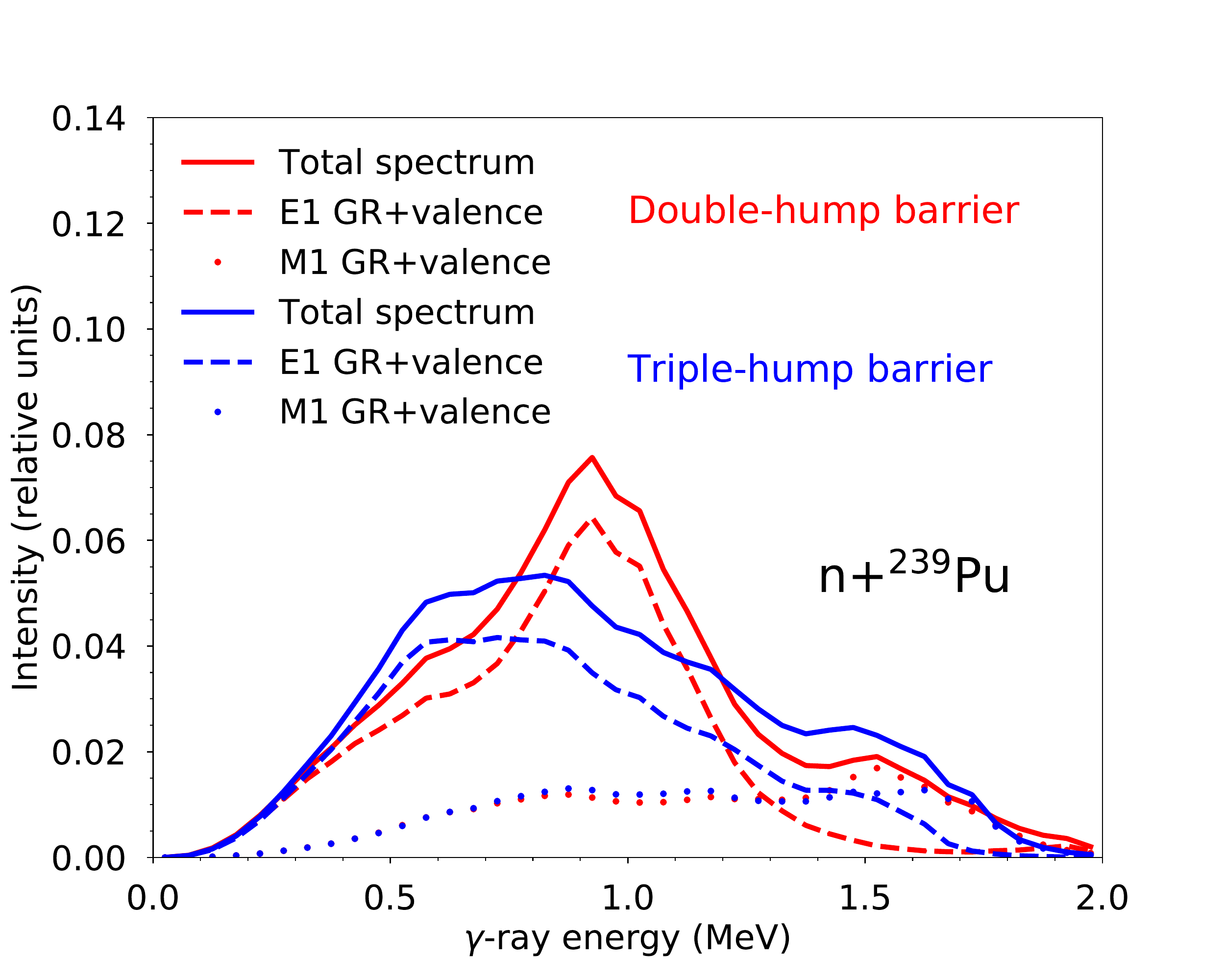}}
\caption{\label{fig:pre-PFGS}Pre-fission \gray~spectrum calculated in the assumptions of double-hump (red lines) and triple-hump (blue lines) barrier.}
\end{figure}

The 1$^+$ states can make E1 transitions to 0$^-$, 1$^-$ and 2$^-$ final states and M1 transitions to 0$^+$, 1$^+$ and 2$^+$ states. Although the M1 radiation widths are about an order of magnitude weaker than those of the E1 transitions in this energy range, we see from Figs.~\ref{fig:fp-pu240-secondary} and~\ref{fig:fp-pu240-tertiary} that the fission probability of final states is considerably higher, especially for 0$^+$ and 2$^+$ final states. Therefore the M1 transitions make a significant contribution to \Ggf, especially at higher \gray~energies, or equivalently at lower residual excitation energies. This is shown in Fig.~\ref{fig:pre-PFGS} where the plotted pre-fission \gray~spectrum is calculated in the double-hump model (red lines) and in the triple-hump model (blue lines). The fission probabilities of the secondary and tertiary well vibrational models are qualitatively similar, but there are considerable quantitative differences, especially in the important 1$^-$ states. For the secondary well model we calculate \Ggf~to be 1.76 meV; the contribution to this from M1 transitions is 0.5 meV.  The mean pre-fission \g~energy \egf\ is 0.97 MeV. In the tertiary well model we find \Ggf= 1.68 meV and \egf= 0.95 MeV. These width and mean energy values are very close to those of the secondary well vibration model even though there is a considerable difference in the shape of the spectra.

We referred in the Introduction to the existence of the \gray\ decay through the second well. The intermediate resonance spacing in the fission cross-section of $^{239}$Pu, interpreted as the class-II spacing $D_{II}(1^+)$, is about 500 eV~\cite{James:1969,Paya:1969}. We use our QPVR model, summarized in Section~\ref{sec:formalism-LD}, using Nilsson orbits for an assumed second well deformation of 0.6, pairing gap parameters similar to those in the primary well and a rotational band constant of 0.0035 MeV  to calculate the level spacing as a function of excitation energy in the second well. We find the observed intermediate resonance spacing consistent with a secondary well ground-state value of about 2.9 MeV. With this level density model and our radiation model described in Section II.C we have calculated the class-II radiation width at this class-II excitation energy to be 10.3 meV. The probability for isomeric fission, i.e., fission following a \g\ cascade in the second well populating the shape isomer, is calculated to contribute no more than 0.05\% to the total fission width.

\begin{figure}[ht]
\centerline{\includegraphics[width=\columnwidth]{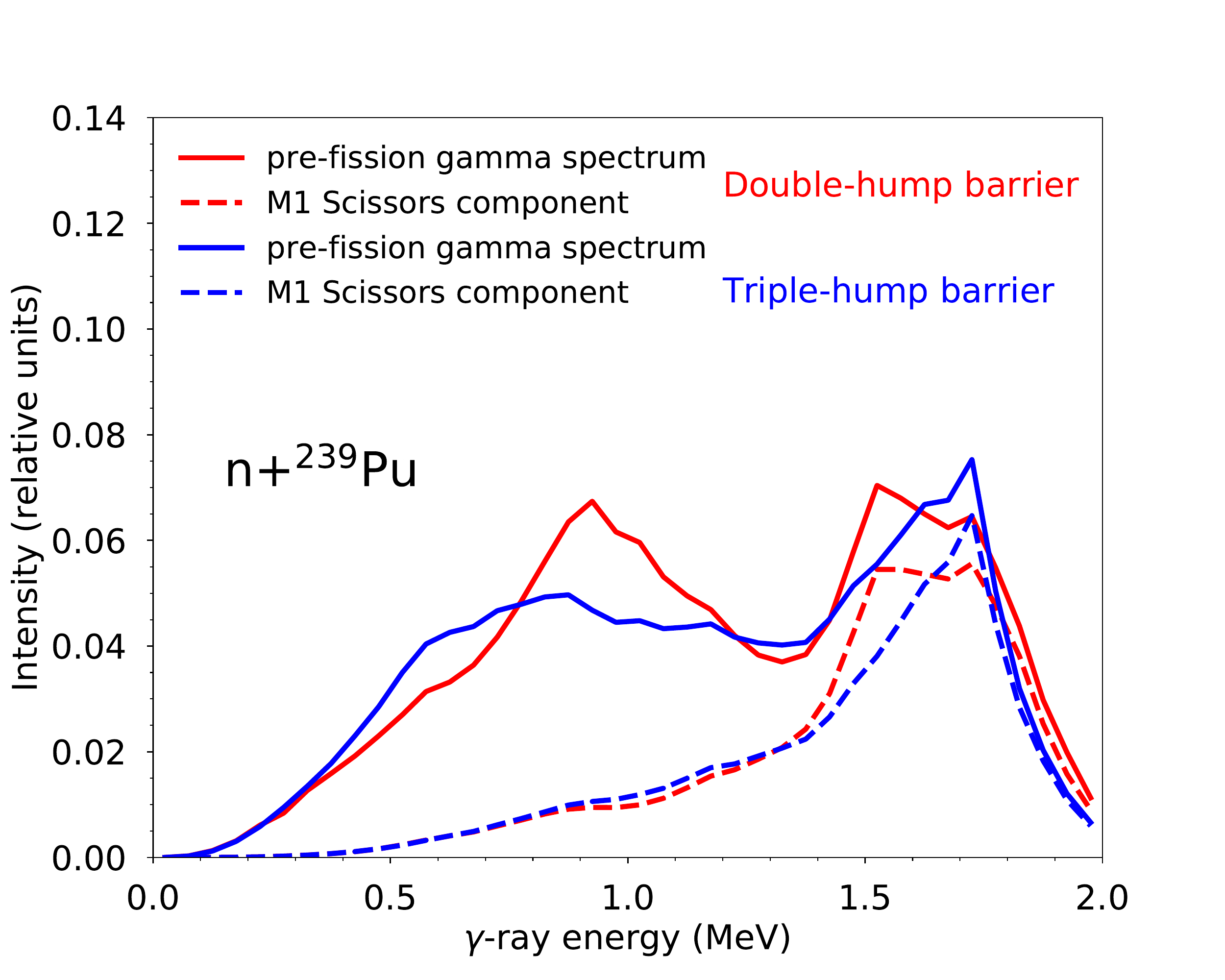}}
\caption{\label{fig:scissors-PFGS}Pre-fission \gray~spectrum if a M1 scissors component is fully added to the \gray~strength function.}
\end{figure}

If the scissors mode, with parameters of the speculative model given in Section~\ref{sec:formalism-Gg}, is included, we calculate that there is an additional contribution of 0.84 meV. In order to retain a total radiation width of 38.5 meV the strength of the standard radiation model described in Section~\ref{sec:formalism-Gg} has to be reduced by 28\%. This model gives \Ggf\ = 2.29 meV (2.17 meV in the tertiary model) and a mean pre-fission \g~energy \egf\ = 1.20 MeV. The pre-fission \g~spectrum and the M1 scissors mode contribution are shown in Fig.~\ref{fig:scissors-PFGS} for the secondary well vibration model (red lines) and for the tertiary well model (blue lines).

\subsection{0$^+$ resonances of $^{239}$Pu+$n$}

In the tertiary well model the total radiation width of spin 0$^+$ resonances is calculated to be 37.9 meV and \Ggf\ = 2.08 meV. The mean pre-fission \g~energy is \egf\ = 0.92 MeV. Nearly all the pre-fission \g~width is through E1 transitions to 1$^-$ final states, and less than 1\% of the width is through M1 transitions. In the ``full-strength'' scissors mode model \Ggf\ = 1.5 meV. There is very little contribution to this from the scissors mode because the 1$^+$ final states have very low fission probability in the region of the scissors mode resonance (see Fig.~\ref{fig:fp-pu240-secondary}).

\subsection{3$^-$ resonances of $^{235}$U+$n$}

In principle these are the most favored resonances of $^{235}$U to show the \ngf~effect. E1 transitions are allowed to 2$^+$ , 3$^+$ and 4$^+$ states and M1 to 2$^-$, 3$^-$ and 4$^-$ states. The fission probabilities for a potential barrier and transition state model that fairly well reproduces the $(d,pf)$ data of Back \etal\cite{Back:1971} are shown in Fig.~\ref{fig:u235-dpf}. The results are \Ggf\ = 2.3 meV and a mean primary gamma-ray energy \egf\ = 0.9 MeV.

\subsection{4$^-$ resonances of $^{235}$U+$n$}
 
\begin{figure}[ht]
\centerline{\includegraphics[width=\columnwidth]{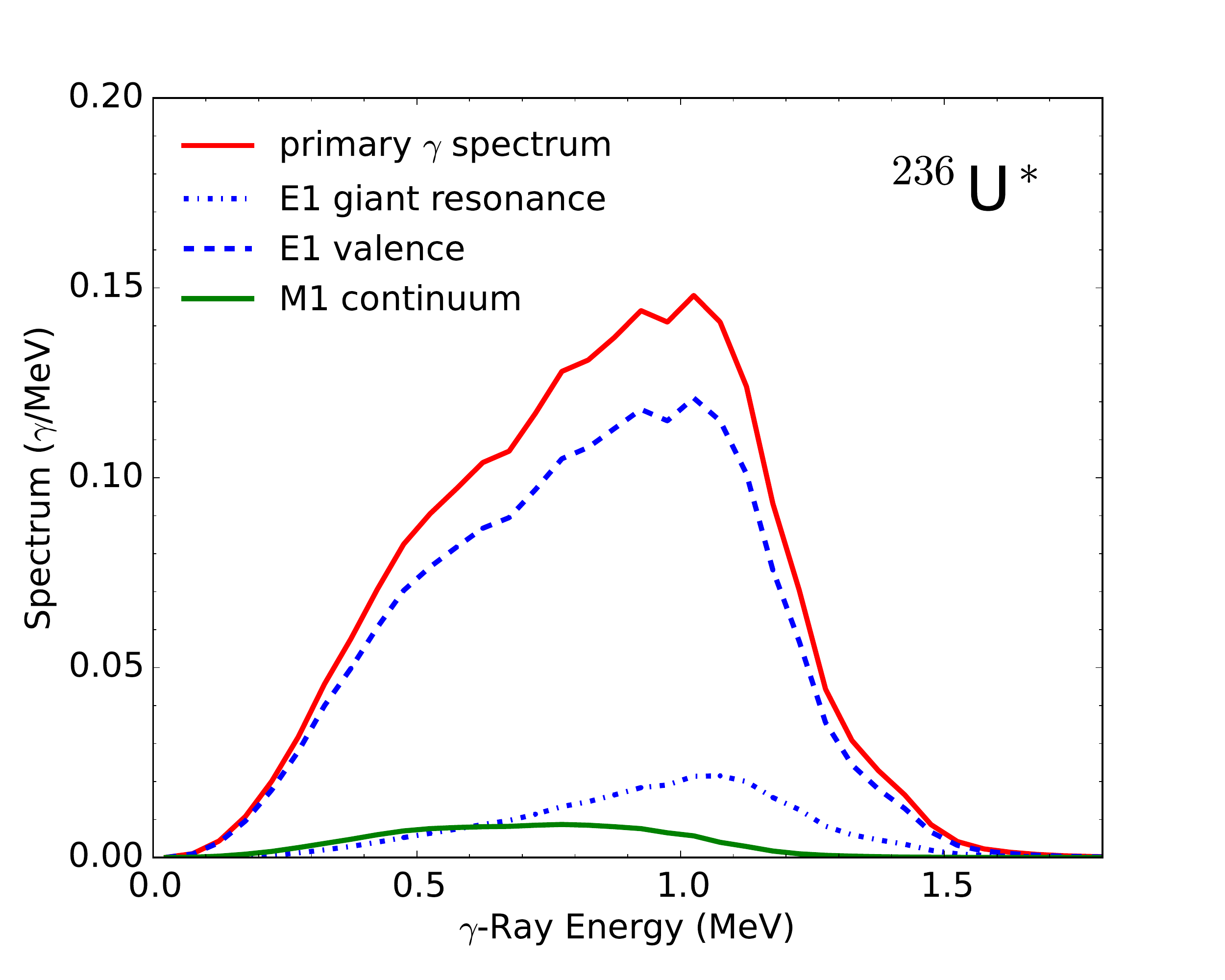}}
\caption{\label{fig:u235-prePFGS}Contributions as function of primary \gray~energy of E1 and M1 transitions of different components of \ngf~reaction in $^{236}$U$^*$.}
\end{figure}

These have E1 transitions to 3$^+$, 4$^+$ and 5$^+$ states and M1 to 3$^-$, 4$^-$ and 5$^-$ states. The results are \Ggf\ = 1.4 meV and a mean primary \gray~energy \egf\ = 0.86 MeV. The spectra of the primary \grays~are shown in Fig.~\ref{fig:u235-prePFGS}. Although the $\gamma f$ width is considerably lower than that of the 3$^-$ resonances, there is considerably better chance of observing the $\gamma f$ process in the 4$^-$ resonances because the average prompt fission width is much lower.

For both spins, M1 transitions contribute only about 5\% to the $^{235}$U $(n,\gamma f)$ reaction. Inclusion of the scissors mode does not much change this, because the fission probability of the final odd-parity states is very low in the scissors mode resonance region (see Fig.~\ref{fig:fp-u236}). An explicit calculation suggests an upper limit of about 0.05 meV. It appears, in fact, that the best candidate for establishing any evidence for the scissors mode is through the $1^+$ resonances in the $^{239}$Pu\ngf\ reaction.

Although there are modeling uncertainties in these estimates, they do not appear to be large. So long as the plausible models for barrier transition states and vibrational resonances fit the experimental fission probability data the variations in predicted widths and mean \g~energies are only a few percent. 

\section{Fast Neutrons} \label{sec:fastresults}

In this Section, we estimate the impact that the \ngf\ process can have on model predictions for both capture and fission channel cross sections in the unresolved resonance and fast energy ranges, above 10 keV and below a few MeV. This incident neutron energy range is particularly important for fast nuclear reactor simulations and other applications.

\begin{figure}[ht]
\centerline{\includegraphics[width=\columnwidth]{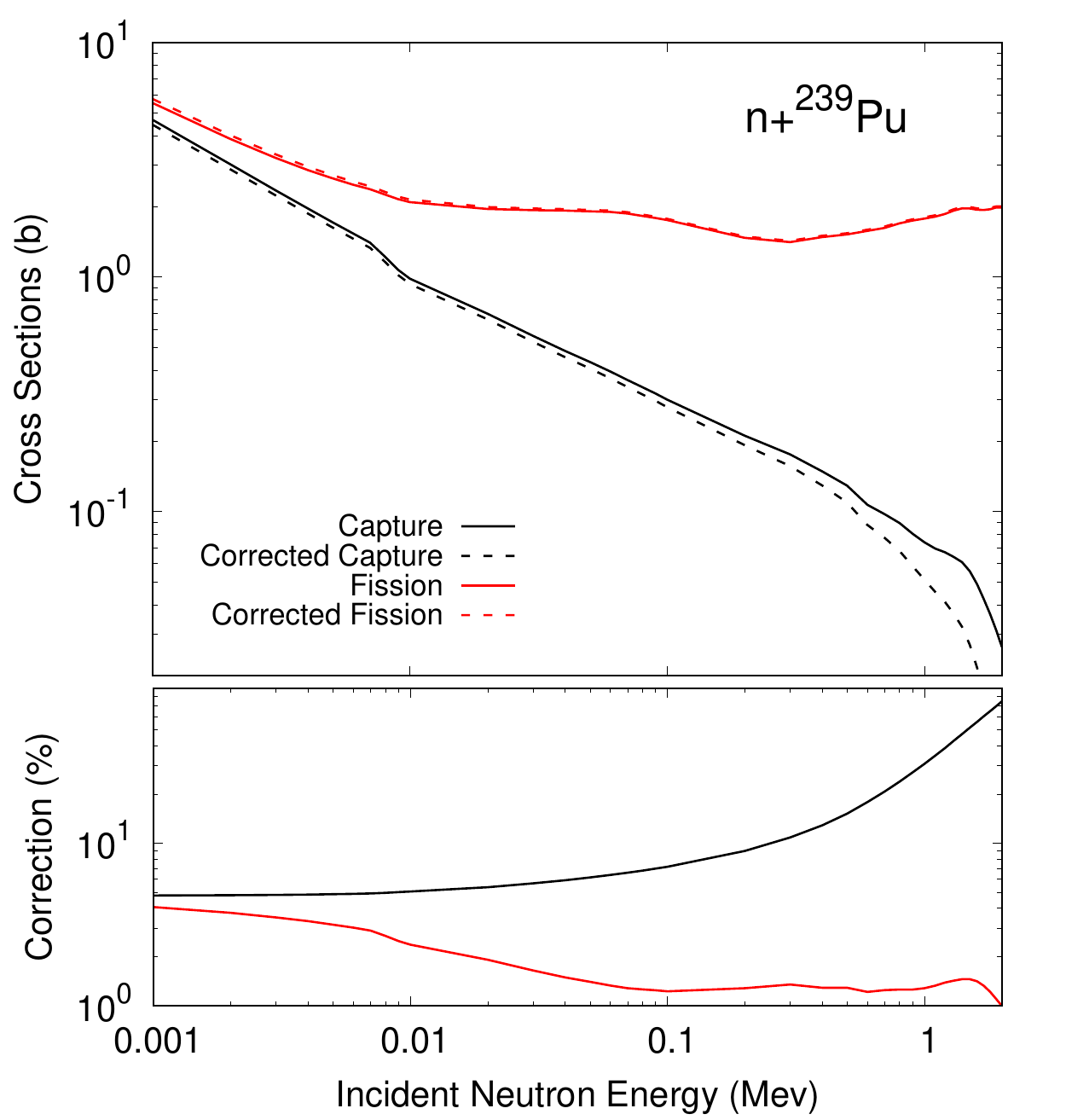}}
\caption{\label{fig:pu239corrections}(top) Neutron-induced capture and fission cross sections on $^{239}$Pu calculated without (solid lines) and with (dashed lines) correcting for the \ngf\ process. (bottom) Magnitude in percent of the cross section corrections caused by the treatment of the \ngf\ process on both capture and fission channels.}
\end{figure}

Figure~\ref{fig:pu239corrections} shows the estimated $(n,\gamma f)$ corrections relative to the neutron-induced fission and capture for $^{240}$Pu$^{*}$ between 1~keV to 2~MeV, calculated in the double-hump barrier model with no M1 scissors mode. The compound nucleus capture cross section is significantly depleted through the \ngf\ process starting from 5\% correction at $E_n=1$~keV to reach 18\% at $E_n=600$~keV where capture is still sizable ($\sigma_\gamma\approx$~100~mb). On the other hand, the calculated fission cross section increases by 1 to 3\% in the same energy range. This correction is comparable in magnitude to the evaluated uncertainties in this energy range~\cite{Talou:2011b}.
 
\begin{figure}[ht]
\centerline{\includegraphics[width=\columnwidth] {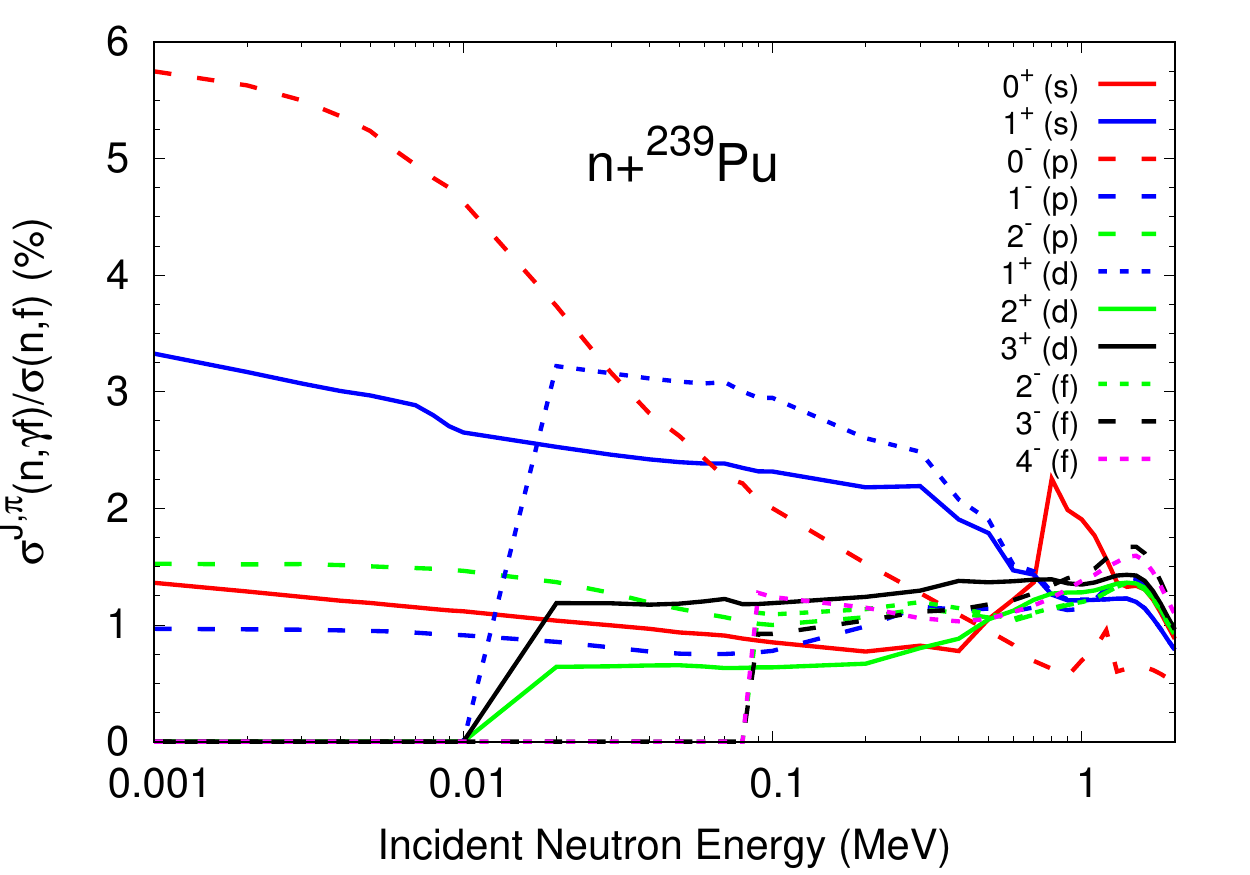} } 
\caption{\label{fig:pu239correctionsF}Exploring Fig.~\ref{fig:pu239corrections} in terms of $l$-wave contributions ($s:J^\pi=0^+,1^+$, $p:J^\pi=0^-,1^-,2^-$, $d:J^\pi=1^+,2^+,3^+$, $f:J^\pi=2^-,3^-,4^-$) relatively here to the neutron-induced fission cross section. $d$ and $f$ wave thresholds, as per arbitrary centrifugal penetrability cutoff, stand at $20$~keV and $100$~keV respectively.}
\end{figure}

\begin{figure}[ht]
\centerline{\includegraphics[width=\columnwidth]{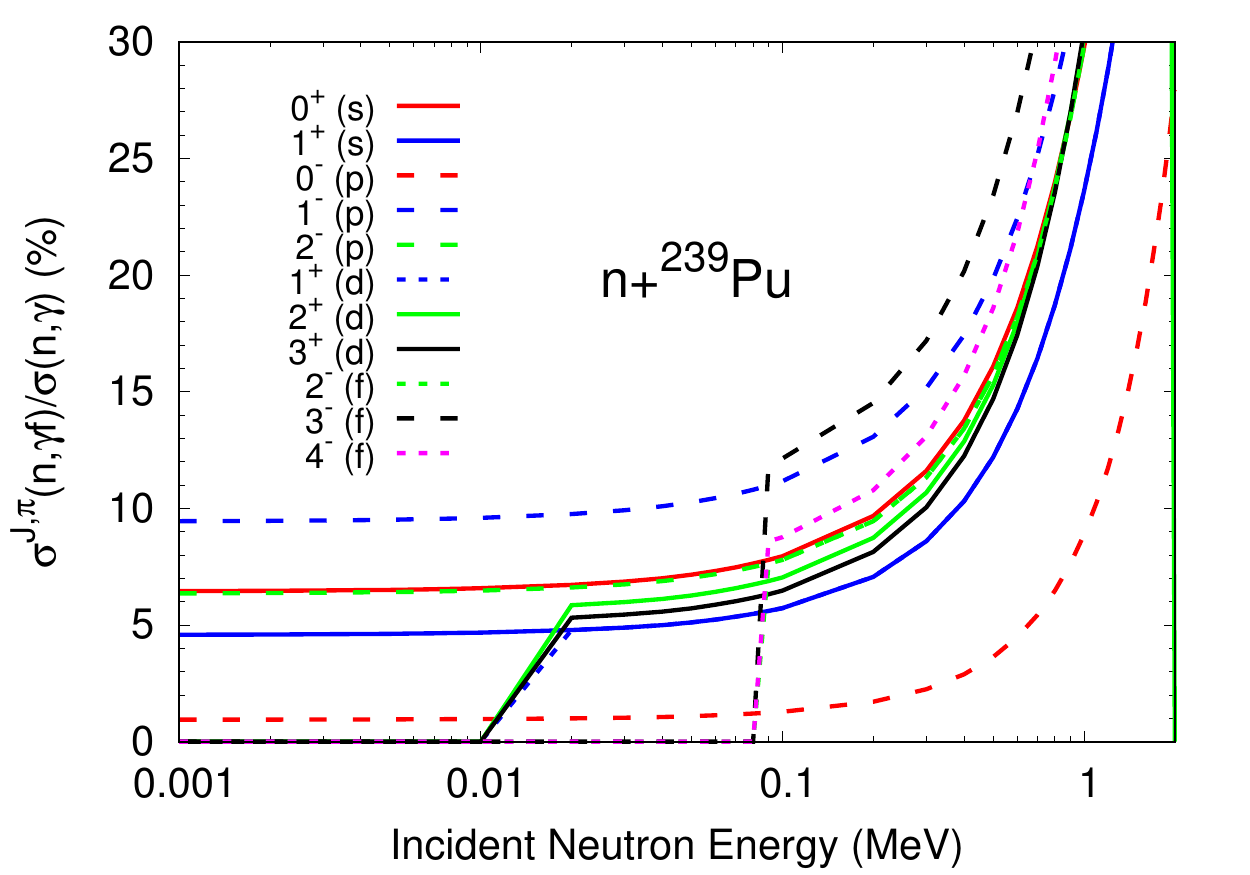}} 
\caption{\label{fig:pu239correctionsG} Same as Fig.~\ref{fig:pu239correctionsF} but for the neutron-induced capture cross section.}
\end{figure}

Figures~\ref{fig:pu239correctionsF} and ~\ref{fig:pu239correctionsG} show the $l$-wave decomposition of the correction factors plotted in Fig.~\ref{fig:pu239corrections} for the fission and capture cross sections, respectively. Figure~\ref{fig:pu239correctionsF} shows that in the resolved resonance region only $s$-waves contributes while in the keV region, $p$, $d$ and $f$ partial waves become important. Since $\Gamma_{\gamma f}$ is rather insensitive to spin and parity at a given neutron energy, the correction is more pronounced for states with small fission probabilities, i.e., for the $0^-$ and $1^+$ states below 100 keV. The fission probabilities for those two states, as shown in Fig.~\ref{fig:fp-pu240-secondary}, are only slowly rising due the absence of $0^-$ and $1^+$ collective states in the transition spectrum (see Section~\ref{sec:formalism-fission} and Table~\ref{tab:collective}). Contributions from $d$ and $f$ partial waves come into play above 20 and 100 keV, respectively.

The $l$-wave decomposition for the capture cross section correction factors plotted in Fig.~\ref{fig:pu239correctionsG} shows that at 1 keV, the $0^-$ and the $1^-$ partial capture relative corrections range from 1\% to 9\%. When the neutron energy exceeds about 300 keV, the observed corrections rise exponentially, following the sharp decrease of the capture cross section in this energy range. 

\begin{figure}[ht]
\centerline{\includegraphics[width=\columnwidth]{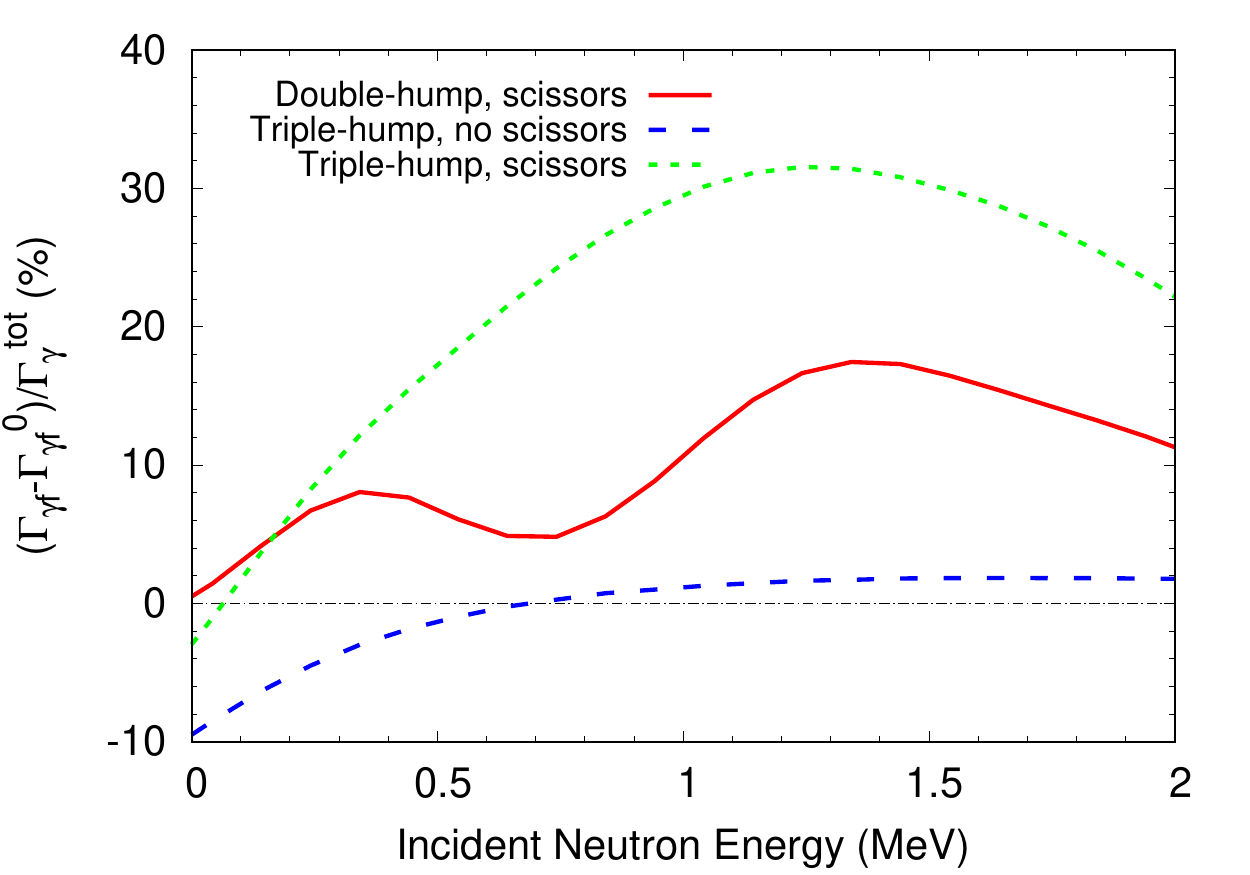}} 
\caption{\label{fig:fast-hypotheses}Relative differences (in percent) between the calculated \Ggf\ using different model assumptions (double vs triple-hump, and M1 scissors mode included or not), with respect to the default \Ggf$^0$ calculated using the double-hump fission barrier model with no M1 scissors mode.}
\end{figure}

There are, of course, modeling uncertainties in these estimates of the corrections. The uncertainties will arise to some degree from the modeling of the barriers. Such uncertainty will be limited by the fact that the barrier modeling will have to reproduce the experimental evidence on fission probability. A greater uncertainty will arise from the modeling of the radiation strength function, in particular, by the strength of the M1 scissors mode. The \ngf\ reaction through the $1^+$ resonances is most affected by this. $J^\pi$=$1^+$ resonances can be excited by $s$-waves and $d$-waves in $^{239}$Pu+$n$, but only by $f$-waves in $^{235}$U+$n$. The effect of different model assumptions on the correction for $J^\pi$=$1^+$ is shown in Fig.~\ref{fig:fast-hypotheses}.
      
It is worth noting that such corrections on capture and fission reaction rates in the fast neutron range would have a significant impact on neutron reactor simulations, for instance. Most theoretical calculations of cross sections do not take those corrections into account properly, thereby adjusting incorrect model parameters to reproduce the experimental data. 

\section{Discussion} \label{sec:discussion}

Fairly convincing qualitative evidence, in terms of the trends of mean \g\ energy and neutron emission as a function of fission width, has been presented by Trochon~\cite{Trochon:1980}, Shackleton \etal\cite{Shackleton:1972} and Frehaut and Shackleton~\cite{Frehaut:1974} for the existence of the \ngf\ reaction. The quantitative results are much more uncertain, however. Trochon gives the result \egf\ = 1.05$\pm$0.05 MeV and \Ggf.\egf\ = 4600$\pm$300 eV$^2$ for the 1$^+$ resonances of $^{239}$Pu. The latter value is two to three times our estimate. If correct, it would require a much softer primary \g\ spectrum than current models allow. Complementary and more accurate measurements would be valuable to test our current models of the fission barrier and capture gamma models. Figures~\ref{fig:pre-PFGS}-\ref{fig:u235-prePFGS} suggest that measurement of the \g\ spectrum associated with fission in resonances with very small measured fission width could reveal the existence of the gamma ray preceding fission and give a direct measurement of its average energy. Accurate measurement of the excess \g\ energy associated with fission in the weak fission resonances, and its ratio to the overall fission \g\ strength from fission products, obtained from the broad fission resonances, could also lead to an accurate estimate of \Ggf, which is expected to remain constant from resonance to resonance.

Likely resonances for making these measurements are presented in Table~\ref{tab:res}. The resonance parameters are from ENDF/B-VII.1~\cite{ENDFB7} and from the Atlas of resonances~\cite{Mughabghab:2006}. For $^{240}$Pu$^*$, there appears to be no 0$^+$ resonances below about 100 eV nearly narrow enough to show an observable \ngf~effect. For $^{236}$U$^*$, it is unlikely that the weakest $3^-$ resonances shown in Table~\ref{tab:res} have a small enough fission width to reveal the primary \grays\ in a spectrum measurement, but increases in total \g\ energy and reduction in \nubar\ may be observable. Finally, although the 4$^-$ resonances appear to have fission widths small enough for the expected \ngf~effects to be observable, they have very small neutron widths or are merged into the wings of much broader resonances, which will make observation much more difficult.

\begin{center}
\begin{table}[ht]
\def\arraystretch{1.25}
\begin{tabular}{cccccc}
\hline
Target & $J^\pi$ & $E_{\rm res}$ (eV) & \Gf (meV)\\
\hline
$^{239}$Pu & $1^+$ & 27.29 & 2.8 \\
& & 35.49 & 3.5 \\
& & 41.46 & 6.4 \\
& & 44.53 & 4.4 \\ 
& & 50.14 & 5.0 \\
& & 82.77 & 5.2 \\
\hline
$^{235}$U & $3^-$ & 2.035 & 10.1 \\
& & 39.13 & 10.5 \\
& & 41.86 & 12.6 \\
& & 43.38 & 17.2 \\
\hline
$^{235}$U & $4^-$ & 4.85 & 5.1 \\
& & 6.39 & 12.4 \\
& & 11.67 & 6.3 \\
& & 18.99 & 4.0 \\
& & 23.42 & 11.0 \\
& & 42.70 & 3.2 \\
& & 49.43 & 12.5 \\
& & 51.62 & 1.2 \\
& & 64.30 & 4.4 \\ 
& & 82.63 & 12.8 \\
& & 94.07 & 8.5 \\
\hline
\end{tabular}
\caption{\label{tab:res}Parameters of the resonances that appear below 100 eV in the $n+^{235}$U and $n+^{239}$Pu reactions, and which are the most likely to exhibit an observable \ngf~effect. These values are taken from the 2006 version of the Atlas~\cite{Mughabghab:2006}.}
\end{table}
\end{center}

Experimental efforts aimed at measuring the pre-fission \grays\ emitted in the \ngf\ reaction have to overcome an important hurdle: most prompt fission \grays\ come from the decay of fission fragments following or in competition with prompt neutron emission. On average, 8$-$9 prompt \grays\ are emitted in the slow-neutron induced fission reactions on  $^{235}$U and $^{239}$Pu, making it very difficult to detect the additional lone \gray\ that would be emitted in the \ngf\ process. Recent calculations~\cite{Stetcu:2014} of the prompt fission \gray\ spectrum make it possible to combine the present work and a calculation of the rest of the emitted prompt \grays\ to obtain \g\ spectra on and off resonances. By inferring the ratios of those spectra, one can then predict a signal for those pre-fission \grays\ and compare to experimental data.

\begin{figure}[ht]
\centerline{\includegraphics[width=\columnwidth]{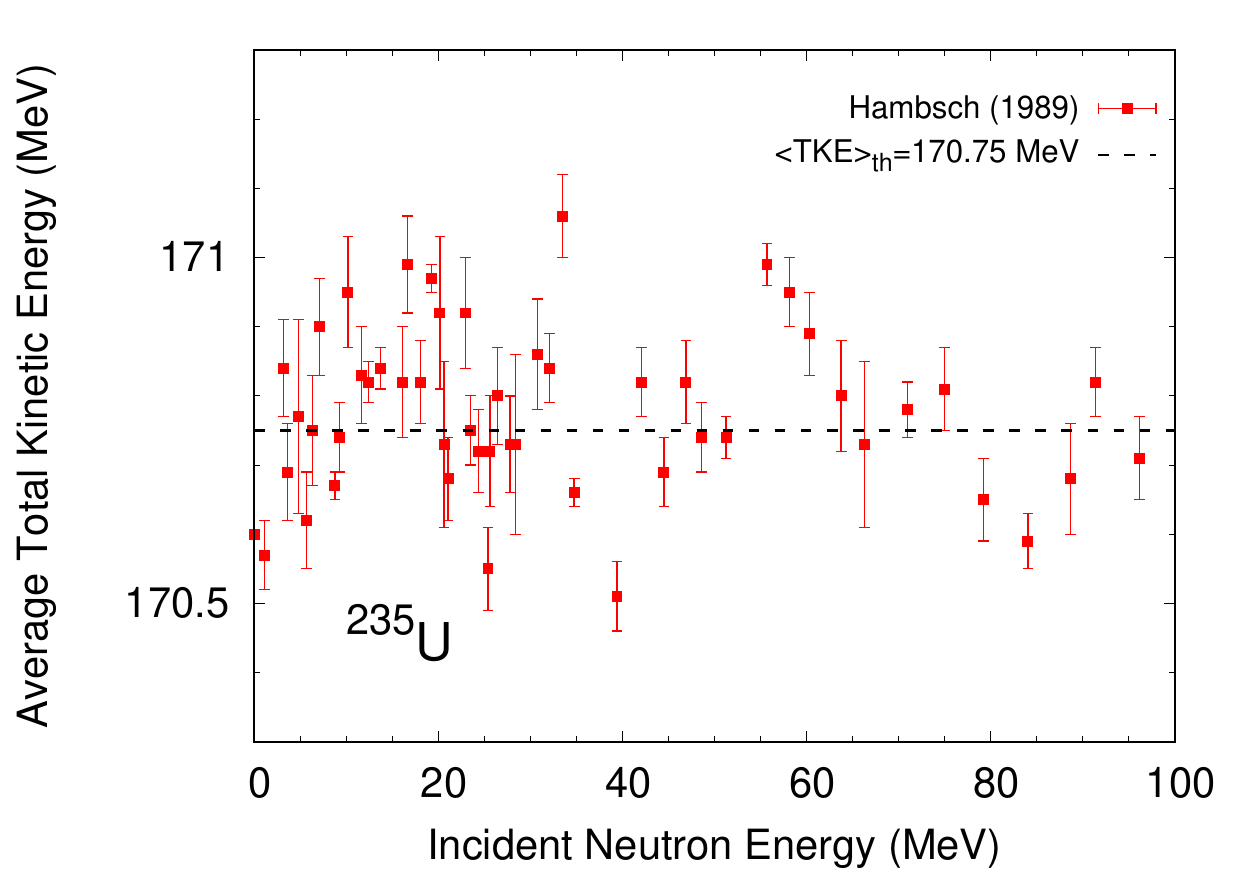}}
\caption{\label{fig:TKEfluct_U235} The average total kinetic energy $\langle$TKE$\rangle$ of the fission fragments produced in the neutron-induced fission reaction of $^{235}$U have been observed by Hambsch \etal~\cite{Hambsch:1989}.}
\end{figure}

\begin{figure}[ht]
\centerline{\includegraphics[width=\columnwidth]{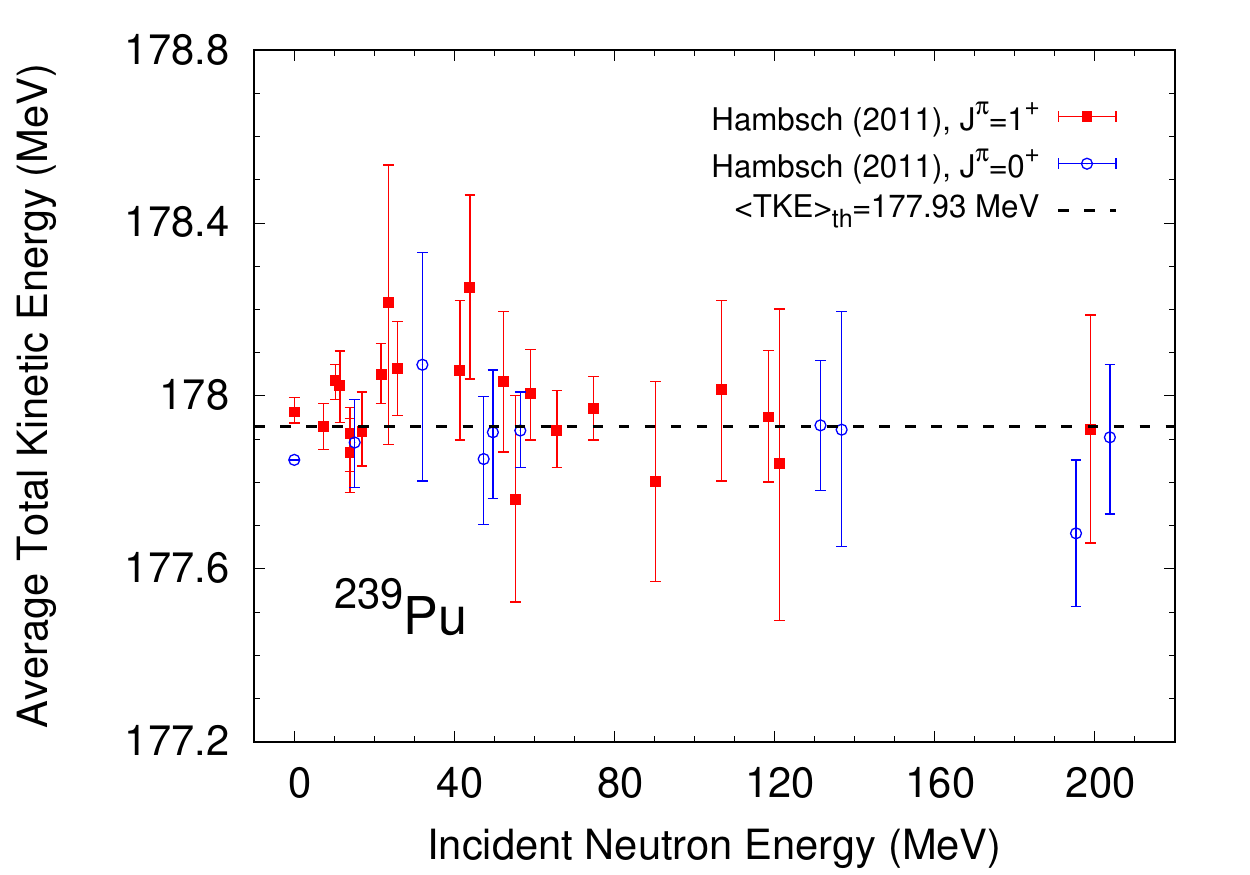}}
\caption{\label{fig:TKEfluct_Pu239} Same as Fig.~\ref{fig:TKEfluct_Pu239} for n+$^{239}$Pu with experimental data from Hambsch \etal~\cite{Hambsch:2012}.}
\end{figure}

An alternative explanation for the fluctuations of \nubar\ in the resonance region has been proposed~\cite{Hambsch:1989,Hambsch:2012}, which merits a discussion. Pre-neutron emission fission fragment yields in mass and total kinetic energy, Y(A,TKE), have a strong impact on the number of prompt fission neutrons emitted. Any fluctuation of those yields as a function of resonance energy would therefore influence \nubar. Yield fluctuations in the resonance region were measured for both $^{235}$U~\cite{Hambsch:1989} and $^{239}$Pu~\cite{Hambsch:2012}. Figure~\ref{fig:TKEfluct_U235} shows the observed fluctuations of the average total kinetic energy $\langle$TKE$\rangle$ up to 100 eV incident neutron energy of $^{235}$U. Somewhat smaller fluctuations were also observed recently for $^{239}$Pu~\cite{Hambsch:2012} and are shown in Fig.~\ref{fig:TKEfluct_Pu239}.

\begin{figure}[ht]
\centerline{\includegraphics[width=\columnwidth]{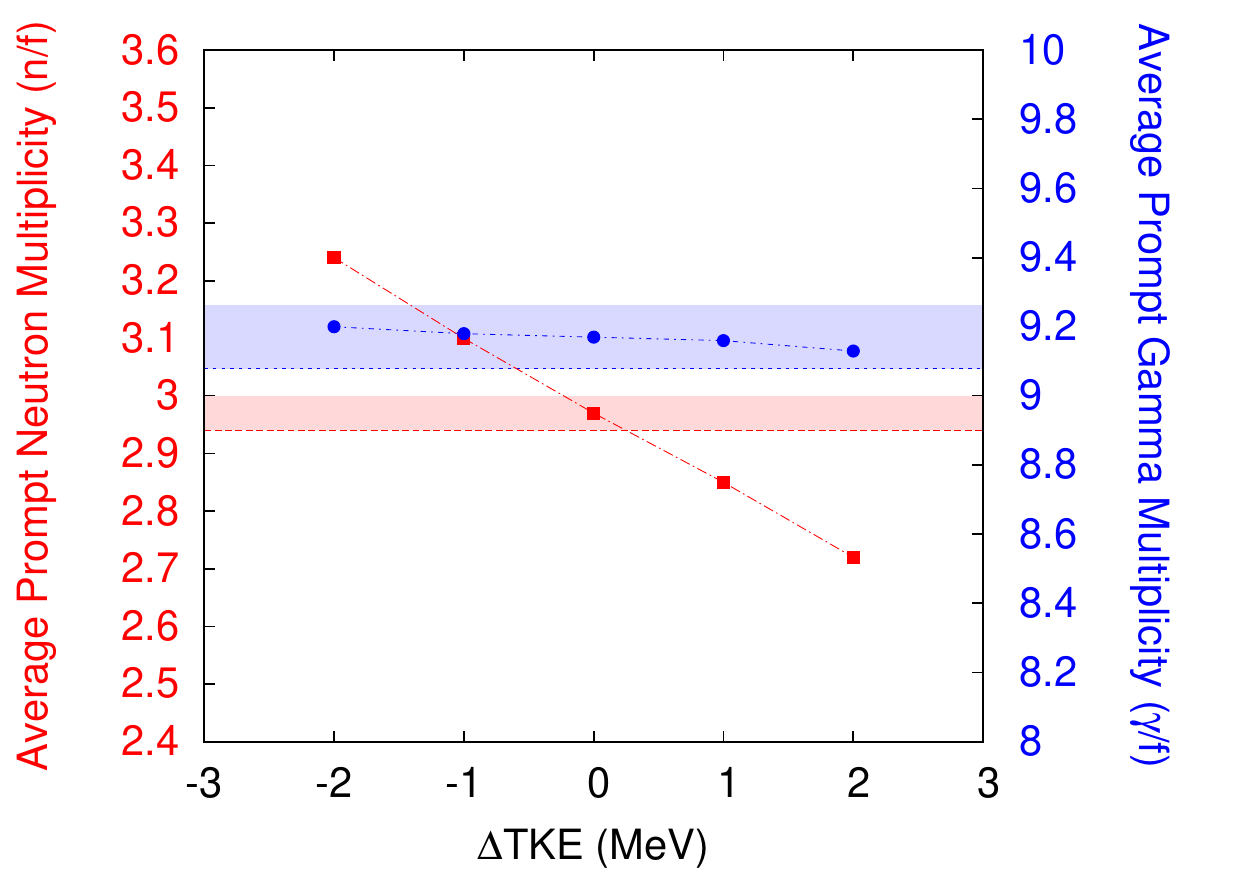}}
\caption{\label{fig:TKE} Influence of the average total kinetic energy of the fission fragments in the neutron-induced fission reaction on $^{239}$Pu on the calculated average neutron and photon multiplicities.}
\end{figure}

To study the impact of those fluctuations on the prompt neutron and photon multiplicities, we performed Monte Carlo Hauser-Feshbach sensitivity calculations of the de-excitation of the fission fragments by varying TKE around its mean value in the case of thermal-neutron-induced fission of $^{239}$Pu. The results are shown in Fig.~\ref{fig:TKE}. As expected, the neutron multiplicity (in red) is strongly impacted by a change in TKE with a calculated slope of $\partial \overline{\nu}_n / \partial TKE \sim -0.13 n/f$/MeV. On the other hand, the average prompt $\gamma$ multiplicity $\overline{N}_\gamma$ (in blue) barely changes with drastic (up to 4 MeV) changes in TKE. Those results indicate that such fluctuations in TKE would not explain the strong fluctuations observed in the average total \gray~energy $\langle E_\gamma^{tot}\rangle$. However, it is also clear that any fluctuation in Y(A,TKE) between resonances would certainly impact \nubar\, indicating that only a complete and correlated study of prompt fission neutron and \gray\ multiplicities in the resonance region can provide the data needed to accurately account for \nubar\ fluctuations.

\section{Conclusion} \label{sec:conclusion}

In this paper, we have revisited an old problem with modern theoretical tools. The \ngf\ process postulated theoretically over 50 years ago remains of great interest to our fundamental understanding of the fission process as well as to nuclear data evaluations of prompt fission data, e.g., neutron and \g\ multiplicities, highly relevant for ongoing R\&D work in advanced nuclear energy systems.

Recent experimental data obtained with the DANCE calorimeter at Los Alamos prompted renewed theoretical calculations with modern tools to compute the fission cross sections in a coupled double-humped barrier configuration. By decomposing the calculated fission cross sections into its spin and parity components, a pre-fission \gray\ spectrum could be inferred. This spectrum is dominated at the lowest energies by E1 transitions, while the high-energy tail of the spectrum M1 transitions in the case of $n+^{239}$Pu. At first, this is surprising since the M1 strength function is about an order of magnitude smaller than the E1 strength. However, fission probabilities of $0^+$ and $2^+$ states, which can only be reached through M1 decay from $s$-wave formed compound nucleus states, dominate all other spin-parity state fission probabilities, hence compensating for the weakness of the M1 strength.

We have also investigated the role that a postulated M1 ``scissors" mode would have on our results. Its main impact would be to increase the pre-fission \g\ spectrum above about 1.2 MeV (see Figs.~\ref{fig:pre-PFGS} and~\ref{fig:scissors-PFGS}). Another change that might have some impact is the 0.2~MeV increase in the mean pre-fission \g\ energy when overlapping an individual fission probability energy threshold. This work also demonstrates strong sensitivity of computed \Ggf\ values to fission barrier and M1 decay mode hypotheses and to input nuclear structure data. In view of the above,  an uncertainty of the order of 20\% has to be applied on to the present \Ggf\ numerical  values. The present study also shows that the importance of the \ngf\ contribution to the calculation of the neutron-induced fission and capture cross sections, even in the fast energy range. 

Our model is obviously an oversimplification of the very complex fission process. However, its consistent treatment of fission calculations with and without the \ngf\ contribution provides a reasonable estimate of the impact of this physical effect, which should be taken into account in modern fission cross section calculations.

Measuring the pre-fission \gray\ spectrum is obviously a difficult task. However, by measuring ratios of PFGS on and off resonances, the large background of prompt \grays\ coming from the decay of the fission fragments can be somewhat removed. 

\section*{Acknowledgements}

This work was performed at Los Alamos National Laboratory, under the auspices of the National Nuclear Security Administration of the U.S. Department of Energy at Los Alamos National Laboratory under Contract No. DE-AC52-06NA25396. P.T. would like to acknowledge some stimulating discussions with Y.~Danon, F.-J.~Hambsch, A.~G\"o\"ok and M.~Pigni about this topic.

\newpage
\bibliographystyle{unsrt}

\begin{thebibliography}{99}

\bibitem{Stavinsky:1965} V.~Stavinsky and M.O.~Shaker, Nucl. Phys. {\bf 62}, 667 (1965).
\bibitem{Lynn:1965} J.E.~Lynn, Phys. Lett. {\bf 18}, 31 (1965).
\bibitem{Bowman:1967} C.~D.~Bowman, G.~F.~Auchampauch, W.~F.~Stubbins, T.~E.~Young, F.~B.~Simpson, and M.~S.~Moore, Phys. Rev. Lett. {\bf 18}, 15 (1967).
\bibitem{Vandenbosch:1967} R.~Vandenbosch, Nucl. Phys. {\bf A 101}, 460 (1967).
\bibitem{Shackleton:1972} D.~Shackleton, J.~Trochon, J.~Frehaut, and M.~Le~Bars, Phys. Lett. {\bf 42B}, 344 (1972).
\bibitem{Ryabov:1973} Yu.~Ryabov, J.~Trochon, D.~Shackleton, and J.~Frehaut, Nucl. Phys. {\bf A216}, 395 (1973).
\bibitem{Frehaut:1974} J.~Frehaut and D.~Shackleton, in proceedings of the Symposium on ``Physics and Chemistry of Fission," Rochester, NY, 13-17 Aug. 1973, International Atomic Energy Agency, Vienna, Vol. II, p.201 (1974).
\bibitem{Howe:1976} R.E.~Howe, T.W.~Phillips, and C.D.~Bowman, Phys. Rev. C {\bf 13}, 195 (1976).
\bibitem{Moore:1978} M.S.~Moore, J.D.~Moses, G.A.~Keyworth, J.W.T.~Dabbs, and N.W.~Hill, Phys. Rev. C {\bf 18}, 1328 (1978).
\bibitem{Shcherbakov:1990} O.A.~Shcherbakov, Fiz. Elemn. Chastits At. Yadra {\bf 21}, 419 (1990).
\bibitem{Mughabghab:2006} S.F.~Mughabghab, ``Atlas of Neutron Resonances," 5th Edition, Elsevier, 2006.
\bibitem{Trochon:1970} J.~Trochon, H.~Derrien, B.~Lucas, and A.~Michaudon, in Proc. of Second Int. Conf. on Nuclear Data for Reactors, Helsinki, June 15-19, 1970, IAEA, Vol. I, p. 495, STI/PUB/259 (1970).
\bibitem{Trochon:1980} J.~Trochon, Proceedings of the Conference {\it Physics and Chemistry of Fission}, J\"u lich, 14-18 May 1979, IAEA-SM-241/A5 (1980).
\bibitem{Fort:1988} E.~Fort, J.~Frehaut, H.~Tellier and P.~Long, Nucl. Sci. Eng. {\bf 99}, 375 (1988).
\bibitem{ENDFB7} M.B.~Chadwick \etal, Nuclear Data Sheets {\bf 107}, 2931 (2006).
\bibitem{WPEC34} ``Coordinated Evaluation of Pu-239 in the Resonance Region," OECD NEA/NSC/WPEC/DOC(2014)447, C.~De~Saint~Jean, R.D.~McKnight, et al. (2014).
\bibitem{Carlson:2009} A.D.~Carlson \etal, Nucl. Data Sheets {\bf 110}, 3215 (2009).
\bibitem{Weston:1974} L.W.~Weston and J.H.~Todd, Phys. Rev. C{\bf 10}, 1402 (1974).
\bibitem{Gwin:1984} R.~Gwin, R.R.~Spencer and R.W.~Ingle, Nucl. Sci. Eng. {\bf 87}, 381 (1984).
\bibitem{Capote:2016} R.~Capote \etal, Nuclear Data Sheets {\bf 131}, 1 (2016).
\bibitem{Reed:1973} R.L.~Reed, PhD Thesis, ``Neutron Multiplicity Measurements for Neutron-Induced Fission of $^{233}$U and $^{235}$U,'' Rensselaer Polytechnic Institute, Troy, New York (1973).
\bibitem{Simon:1975} G.~Simon and J.~Frehaut, in Neutron Physics: Proc. of the {\it Third AlI-Union Conference on Neutron Physics,} June 9-13,1975, Kiev (in Russian), Institute of Atomic Information, Moscow, 1976, Part 5, p. 337 (1976).
\bibitem{Hambsch:1989} F.-J.~Hambsch, H.-H.~Knitter, C.~Budtz-J\o rgensen, and J.~P.~Theobald, Nucl. Phys. {\bf A491}, 56 (1989).
\bibitem{Fort:2008} E.~Fort and A.~Courcelle, in proceedings of the International Conference on Nuclear Data for Science \& Technology ND2007, April 22-27, 2007, Nice, France, EDP Sciences (2008).
\bibitem{Hambsch:2017} F.-J.~Hambsch in Proceedings of the Sixth Int. Conf. on Fission and Properties of Neutron-Rich Nuclei, Nov. 6-12, 2016, Florida, USA, {\it to appear in} World Scientific, Eds. J.~Hamilton, A.~V.~Ramayya and P.~Talou (2017).
\bibitem{Guttormsen:2012} M.~Guttormsen, L.A.~Bernstein, A.~B\"urger, A.~G\"orgen, F.~Gunsing, T.W.~Hagen, A.C.~Larsen, T.~Renstr\o m, S.~Siem, M.~Wiedeking, and J.N.~Wilson, Phys. Rev. Lett. {\bf 109}, 162503 (2012).
\bibitem{Guttormsen:2014} M.~Guttormsen, L.A.~Bernstein, A.~G\"orgen, B.~Jurado, S.~Siem, M.~Aiche, Q.~Ducasse, F.~Giacoppo, F.~Gunsing, T.W.~Hagen, A.C.~Larsen, M.~Lebois, B.~Leniau, T.~Renstr\o{}m, S.J.~Rose, T.G.~Tornyi, G.M.~Tveten, M.~Wiedeking, and J.N.~Wilson, Phys. Rev. C {\bf 89}, 014302 (2014).
\bibitem{Ullmann:2014} J.L.~Ullmann and T.~Kawano and T.A.~Bredeweg and A.~Couture and R.~C.~Haight and M.~Jandel and J.M.~O'Donnell and R.S.~Rundberg and D.J.~Vieira and J.B.~Wilhelmy and J.A.~Becker and A.~Chyzh and C.Y.~Wu and B.~Baramsai and G.E.~Mitchell and M.~Krti\v{c}ka, Phys. Rev. C {\bf 89}, 034603 (2014).
\bibitem{Mumpower:2017} M.~R.~Mumpower and T.~Kawano and J.L.~Ullmann and M.~Krti\v{c}ka and T.M.~Sprouse, Phys. Rev. C {\bf 96}, 024612 (2017).
\bibitem{Ullmann:2017} J.L.~Ullmann, T.~Kawano, B.~Baramsai, T.A.~Bredeweg, A.~Couture, R.C.~Haight, M.~Jandel, J.M.~O'Donnell, R.S.~Rundberg, D.J.~Vieira, J.B.~Wilhelmy, M.~Krti\ifmmode \check{c}\else \v{c}\fi{}ka, J.A.~Becker, A.~Chyzh, C.Y.~Wu and G.E.~Mitchell, Phys. Rev. C {\bf 96}, 024627 (2017).
\bibitem{Bouland:2013} O.~Bouland, J.E.~Lynn, and P.~Talou, Phys. Rev. C {\bf 88}, 054612 (2013).
\bibitem{Gustafson:1967} C.~Gustafson, I.L.~Lamm, B.~Nilsson, S.~G.~Nilsson, Ark. Fys. {\bf 36}, 613 (1967).
\bibitem{Kopecky:1990} J.~Kopecky and M.~Uhl, Phys. Rev. C {\bf 41}, 1941 (1990).
\bibitem{Spencer:1986} R.R.~Spencer, J.A.~Harvey, N.~W.~Hill, and L.W.~Weston, Proceedings of the International Conference on {\it Nuclear Data for Basic and Applied Science}, Santa Fe, May 13-17, 1985, Eds. P.G. Young \etal, Vol. 1, p. 581, Gordon and Breach, New York, 1986.
\bibitem{Bjornholm:1980} S.~Bj\o rnholm and J.~E.~Lynn, Rev. Mod. Phys. {\bf 52}, No. 4, 725 (1980).
\bibitem{Cramer:1970} J.D.~Cramer and H.~C.~Britt, Phys. Rev. C {\bf 2}, 2350 (1970).
\bibitem{Singh:2002} B.~Singh, R.~Zywina, and R.B.~Firestone, Nuclear Data Sheets {\bf 97}, 2, Pages 241-592, Third Edition (October 2002), https://doi.org/10.1006/ndsh.2002.0018. 
\bibitem{Ichikawa:2013} T.~Ichikawa, P.~M\"oller, and A.J.~Sierk, Phys. Rev. C {\bf 87}, 054326 (2013).
\bibitem{Back:1971} B.B.~Back, J.~P.~Bondorf, G.A.~Ostroschenko, J.~Pedersen and B.~Rasmussen, Nucl. Phys. {\bf A165}, 449 (1971).
\bibitem{James:1969} G.D.~James, and B.H.~Patrick, Proceedings of the Second IAEA Symposium on {\it Physics and Chemistry of Fission}, Vienna , July 28 - August 1, 1969,  p. 391, IAEA, Vienna, Austria, 1969.
\bibitem{Paya:1969} D.~Paya, J.~Blons, H.~Derrien, and A.~Michaudon, Proceedings of the Second IAEA Symposium on {\it Physics and Chemistry of Fission}, Vienna , July 28 - August 1, 1969,  p. 307, IAEA, Vienna, Austria, 1969.
\bibitem{Talou:2011b} P.~Talou and P.G.~Young and T.~Kawano and M.~E.~Rising and M.~B.~Chadwick, Nuclear Data Sheets {\bf 112}, 3054 (2011).
\bibitem{Hambsch:2012} F.-J.~Hambsch, I.~Ruskov, and L.~Dematt\'{e}, in Theory-1, Proceedings of the Scientific Workshop on Nuclear Fission Dynamics and the Emission of Prompt Neutrons and Gamma Rays, Sep. 27-29, 2010, Sinaia, Romania, JRC/IRMM, EUR 24802 EN, p.41 (2011).
\bibitem{Stetcu:2014} I.~Stetcu, P.~Talou, T.~Kawano, and M.~Jandel, Phys. Rev. C {\bf 90}, 024617 (2014).

\end{thebibliography}

\end{document}